\documentclass[sn-aps]{sn-jnl}% Math and Physical Sciences Reference Style

\usepackage{natbib}
\newcommand\proy {\mbox{\boldmath $P$}}
\newcommand{\convprob  }{ \buildrel{p}\over\longrightarrow}
\newcommand\bv {\mathbf v}
\newcommand\bu {\mathbf u}
\newcommand{\cp}{\overset{P}{\rightarrow}}
\newcommand{\spn}{\rm{span}}
\newcommand\bx {\mathbf x}
\newcommand\balfa {\mbox{\boldmath $\alpha$}}
\newcommand{\convdist}{ \buildrel{D}\over\longrightarrow}
\jyear{2021}%
 \newcommand\bz {\mathbf z}
\newcommand\bX {\mathbf X}
\theoremstyle{thmstyleone}%
\newtheorem{theorem}{Theorem}%  meant for continuous numbers
\newtheorem{proposition}[theorem]{Proposition}% 

\theoremstyle{thmstyletwo}%
\newcommand\bW {\mathbf W}
\theoremstyle{thmstylethree}%
\newcommand\bw {\mathbf w}
\newcommand{\esp}{\mathbb{E}}
\newcommand\bbe {\mbox{\boldmath $\beta$}}

\newcommand\regd{m}

\def\EE{\esp}
\def\var{\hbox{var}}
\def\XX{\mathbf X}
\def\xx{\mathbf x}

\def\PP{\mathrm P}
 
\def\regd{\xi}
\def\regden{\ell}

\def\gn{g_{\widehat\bbe_n,\xx_0}}
\def\gnt{g_{{{\tiny{\widehat\bbe_n}},\xx_0}}}

\def\g0t{g_{{\tiny{\bbe_0}},\xx_0}}
\def\trans{t}

\def\gnl{r}

\begin{document}

\title[Asymptotic results for  prediction after  sufficient dimension reduction]{Asymptotic results for  nonparametric regression estimators after sufficient dimension reduction estimation}


\author*[1,2]{\fnm{Liliana} \sur{Forzani}}\email{lforzani@fiq.unl.edu.ar}

\author[2,3]{\fnm{Daniela} \sur{Rodriguez}}\email{drodrig@dm.uba.ar}
\equalcont{These authors contributed equally to this work.}

\author[2,3]{\fnm{Mariela} \sur{Sued}}\email{msued@dm.uba.ar}
\equalcont{These authors contributed equally to this work.}

\affil*[1]{\orgdiv{Facultad de Ingenier\'ia Qu\'imica}, \orgname{Universidad Nacional del Litoral}, \orgaddress{\city{Santa Fe},  \country{Argentina}}}

\affil[2]{ \orgname{CONICET}, \orgaddress{ \country{Argentina}}}

\affil[3]{\orgdiv{Facultad de Ciencias Exactas y Naturales}, \orgname{Universidad de Buenos Aires}, \orgaddress{ \city{Buenos Aires},  \country{Argentina}}}


\abstract{Prediction, in regression and classification, is one of the main aims in modern data science. When the number of predictors is large, a common first step is to reduce the dimension of the data. Sufficient dimension reduction (SDR) is a well established paradigm of reduction that keeps all the relevant information in the covariates $X$ that is necessary for the prediction of $Y$. In practice, SDR has been successfully used as an exploratory tool for modelling after estimation of the sufficient reduction. Nevertheless, even if the estimated reduction is a consistent estimator of the population, there is no theory that supports this step when non-parametric regression is used in the imputed estimator. In this paper, we show that the asymptotic distribution of the non-parametric regression estimator is the same regardless if the true SDR or its estimator is used. This result allows making inferences, for example, computing confidence intervals for the regression function avoiding the curse of dimensionality.}




\maketitle

%%\unnumbered% uncomment this for unnumbered level heads

\section{Introduction}

The prediction of  a response $Y \in {\mathbb R}$ based on  a random vector  $\XX \in {\mathbb R}^p$ is one of the main goals in modern science and may be achieved estimating the conditional mean function $\esp(Y\mid \XX)$.  When the number of predictors $p$
is large, most methods  include some
type of dimension reduction for $\XX$. 
Sufficient dimension reduction (SDR)  has proven to be a powerful tool for this task, extracting sufficient information 
%from high-dimensional data, 
in   regression and classification settings (see \cite{cook2009regression} and references thereby).
SDR includes a family of methods that seek a reduction  $R(\XX)$, where $R: {\mathbb R}^p \rightarrow {\mathbb R}^d$ with $d < p$,  that capture all the information that $\XX$ contains about $Y$ in the sense that $Y\mid \XX\sim Y\mid R(\XX)$. In particular, this implies that   $\esp(Y\mid \XX)=\esp(Y\mid R(\XX))$. 
%\cite{li2003dimension} developed different methods to characterized the  function $R(\XX)$ which, even may not be a SDR,  satisfies that $\esp(Y\mid \XX)=\esp(Y\mid R(\XX))$,
%without requiring a specified parametric model.

There are three general paradigms for determining and estimating a sufficient reduction. The  first of them is   based on moments or functions of moments
of the conditional distribution of $\XX \mid Y$ (SIR, \cite{li91}; SAVE, \cite{cookweisberg}; pHd, \cite{li92},
PIR, \cite{bura01}; MAVE, \cite{Xia02},  \cite{CookLi},  \cite{CookNi},  \cite{Li05},\cite{Zhu06}; DR, \cite{LiWang}). The second methodology  postulates a  parametric model for the inverse regression of $\XX$ on $Y$. In this model-based approach, minimal sufficient reductions can be  determined from the model itself  (\cite{Cook2007, cook_forzani_2008, Cook:2009aa,  cook_forzani_2009, Cook:2008aa, adragni2009sufficient, Pfeiffer:2012aa, Zhong:2015aa, BuraDuarteForzani2016, Tomassi:2017aa}).
Under this setting, the SDR, $R(\XX)$, is not necessary linear on $\XX$ as in the case of moment based methodology (see \cite{BuraDuarteForzani2016}).
%A defining feature of sufficient reductions obtained via moments is that they are projections
%of the predictor vector on lower dimensional subspaces and  this applies only to continuous predictors. Therefore, sufficient reductions have been by
%default linear and SDR methodology has mostly been linear SDR. The second framework developed to determine and estimate SDR works by specifying a parametric model for the inverse regression of $\XX$ on $Y$. In this model-based approach, minimal sufficient reductions can be  determined from the model itself  (\cite{Cook2007, cook_forzani_2008, Cook:2009aa,  cook_forzani_2009, Cook:2008aa, Adragni4385, Pfeiffer:2012aa, Zhong:2015aa, BuraDuarteForzani2016, Tomassi:2017aa}).
% Under this approach, the SDR, $R(\XX)$, is not necessary linear on $\XX$ as in the case of moment based methodology (see \cite{BuraDuarteForzani2016}).
Finally, if the target is $\esp (Y\mid \XX)$, instead of the whole conditional distribution $Y\mid \XX$, \cite{li2003dimension} developed methods to get linear reductions under mild conditions. In the same lines, recently  \cite{cook2021pls} found that under a non-linear forward model, Partial least square (PLS) reduction gives sufficient linear combinations to estimate $\esp (Y \mid \XX).$

Reducing the dimension of $\XX$ to predict $Y$ is desirable because it makes statistical methods more efficient.
For instance, the performance of nonparametric regression techniques, such as Nadaraya \cite{nadaraya1964estimating}  - Watson \cite{watson1964smooth}, 
depends on the dimension of the predictor. 
Thus, using such  methods on a reduction should result in better performance than if those procedures were applied without reducing. 
%
%This idea is on the top of the methods presented above: estimating the regression function using the  reduced predictors.
Namely,  estimators based on $(\XX_i,Y_i)$ gives rise to estimations with $\sqrt{nh^p}$ rates, while using $(R(\XX_i), Y_i)$, allows to attempt $\sqrt{nh^d_n}$ rate and, in this way, 
the course of dimensionality may be broken.
However, even when the reduction can be characterized in terms of the joint distribution of $(\XX,Y)$, it must be estimated.
In practice, $(\widehat R(\XX_i), Y_i)$ is available to predict where  $\widehat R$ stands for an estimator of the reduction $R$. 
This is the way sufficient dimension reduction has been used by practitioners, to great success. However, no theoretical results are available to quantify the impact of using the estimated reduction in lieu of the true one in this non parametric setting. 
 For instance, in parametric frameworks this problem has been addressed by \cite{kim2020post}. They show that  treating  the estimated sufficient predictors as the true ones  leads to distorted results when performing inference about the finite dimensional parameter of interest. Prediction after reduction is not discussed in their work. The only  asymptotic 
 theoretical results related to prediction  after reduction is the one given by \cite{forzani2019sufficient} when using the model based paradigm to get the reduction. Using a prediction method 
 developed by \cite{adragni2009sufficient} they   prove that 
 when $\XX\mid Y$ follows a generalized linear model.

The main goal of this work is to prove that nonparametric estimators based on estimated reductions perform as well as if the true reduction were used. As we discuss below, and in contrast to the parametric setting described, the imputation of an estimated reduction with better rate than that of the regression method allows to neglect the impact of the imputation.  Up to now,   only  heuristic grounds were presented to justify this fact and here we provide a rigorous  theoretical proof, filling an important gap which remained open up to now.

In Section \ref{est_reg_red} we defined an estimator of $\esp (Y \mid \XX)$ using the estimated reduction. 
In Section \ref{asintotitca} 
 the asymptotic behaviour  of the estimator is given. The uniform consistency 
 is also discussed at the end of this section. In Section \ref{proy-1} we show how the existing estimators of the reduction can be used to fit the  framework developed in this work. 
 We report the results of a simulation study in Section \ref{simu}.
In Section \ref{ejemplo}, using the example considered in  \cite{cook2018introduction, basa2023asymptotic, cook2023libro},
we show how to compute in practice  approximate confidence intervals for the conditional mean.
%
%{\color{red}{ESTO LO PONEMOS?
%  \cite{gyorfi2002distribution} present a complete overview of the developed machinery in nonparametric regression before the deep learning era.  An exhaustive review of the existent results is unfeasible or, at least,  out of the scope of this work. For sake of completeness, inspired in \cite{ziegler2003asymptotic},  we include in the Appendix  the asymptotic behavior of the kernel based nonparametric regression estimator, which are used to prove the main results presented in this work.  
%
%}}

\section{Estimating the regression function after a linear reduction}
\label{est_reg_red}
	Consider  $(\XX,Y)\in \mathbb R^{p+1}$. Assume that there exists $\bbe_0\in \mathbb R^{d\times p}$ such that
\begin{equation}
	\label{lin_red}
	\esp ({Y\mid \XX ) = \esp (Y\mid  \bbe_0\XX}), 
\end{equation}
and therefore, $R(\XX)=\bbe_0\XX$ is  a sufficient reduction for  $\esp (Y\mid \XX )$. 
Let $\bW=\bbe_0\XX\in \mathbb R^d$ and define 
\begin{equation}
	\label{regd}\eta(\xx_0)=\esp(Y\mid \XX=\xx_0)\;,\quad\regd(\bw)=\esp(Y\mid \bW=\bw).
\end{equation}
Thus, for $\bw_0=\bbe_0\xx_0$ with fixed $\xx_0\in \mathbb R^p$,  we have that 
$\eta(\xx_0)=\regd(\bw_0)$. We are interested in estimating $\eta(\xx_0)$  for a given   $\xx_0$. Note that if $\bbe_0$ were known, 
we could  estimate  $\eta(\xx_0)=\regd(\bw_0)$ with
\begin{equation}
	\label{est-lim1}
	\widehat \regd(\bw_0)=\displaystyle\frac{\sum_{i=1}^n  Y_i K\left(\frac{ \bw_0-\bW_i}{h_n}\right)}
	{\sum_{i=1}^n K\left(\frac{ \bw_0-\bW_i}{h_n}\right)}=
	\frac{\sum_{i=1}^n  Y_i K\left(\frac{ \bbe_0\xx_0-\bbe_0\XX_i}{h_n}\right)}
	{\sum_{i=1}^nK\left(\frac{ \bbe_0\xx_0-\bbe_0\XX_i}{h_n}\right)}, \end{equation}
where $h_n>0$. In this case, under the conditions established in Theorem \ref{EL_teo}, it is known that $\widehat\regd (\bw_0)\convprob  \regd(\bw_0)=\eta(\xx_0)$ and  $	\sqrt{ nh_n^d} \{\widehat \regd(\bw_0)-\regd(\bw_0)\}$ converges in distribution to a Gaussian law.
\textcolor{black}{An exhaustive review of these  results is unfeasible or, at least,  out of the scope of this work. For the sake of completeness, inspired by \cite{ziegler2003asymptotic} we include the proof of this result.  }

In practice, $\bbe_0$ is unknown and thus the estimator {\color{black}{$\widehat \regd(\bw_0)$ with $\bw_0=\bbe_0\xx_0$ cannot be computed.}} 
Guided by the plug-in spirit, we input an estimator $\widehat \bbe_n$ of $\bbe_0$ and define 
\begin{equation}
	\label{est-lim}
	\widehat\eta(\xx_0):= \frac{\sum_{i=1}^n  Y_i K\left(\frac{\widehat{\bbe}_n  (\xx_0-\XX_i)}{h_n}\right)}
	{\sum_{i=1}^n K\left(\frac{\widehat{\bbe}_n  (\xx_0-\XX_i)}{h_n}\right)}.
\end{equation}
In this work we establish that, under regularity conditions stated in Theorem \ref{asymptotic}, 
$	\widehat\eta(\xx_0)$ is asymptotically equivalent to $\widehat \regd(\bw_0)$ as far as $\sqrt{n}(\widehat{\bbe}_n-\bbe_0)=O_P(1)$.
Thus, the reduction  drastically improves the rate of convergence of the estimator without losing efficiency when the  estimated reduction is used instead of the true one.

\textcolor{red}{}

 \section{Assumptions and main results} 
 \label{asintotitca}

%  In this section we show that  $\widehat\eta(\xx_0)$, defined in (\ref{est-lim}), is a consistent estimator of $\eta(\xx_0)$ and derive its asymptotic normality. Interestingly, the  asymptotic behavior of the estimator does not depends on that of
% $\widehat{\bbe}_n$ when $n^{1/2}(\widehat\bbe_n-\bbe_0)=O_P(1)$.
 
% At this point we would like to emphasize that this fact  \textcolor{red}{transcends} the sufficient dimension reduction framework presented in this work. We are, indeed, exploring a very general result which mainly states that the imputation of an estimator (like $\widehat{\bbe}_n$) in a procedures with worst rate of convergence when considering a true parameter ($\bbe_0$) remains unaffected by the use of the estimator in lieu of its limit as far as the last one has a better rate of convergence. 
% 
% 

 Let $(\XX_i,Y_i)$, $i\geq 1$,  be independent and identically  distributed copies of $(\XX,Y)$. 
 Consider  $\bW=\bbe_0\XX$ and $\bw_0=\bbe_0\xx_0$, where   $\bbe_0$ and $\xx_0$  are a fixed matrix in $\mathbb R ^{d\times p}$  and a point in $\mathbb R ^{ p}$, respectively. Define 
 \begin{eqnarray}
 	\label{regd}
 	\regd(\bw)&=&\esp(Y\mid \bW=\bw),\nonumber\\
		\regd_2(\bw)&=&\esp(Y^2\mid \bW=\bw),\nonumber\\
	 \sigma^2(\bw)&=&
 	\esp\left(\left\{Y -\regd(\bW)\right\}^2\mid \bW=\bw \right).
 \end{eqnarray}

\subsection{Pointwise consistency and asymptotic normality}

  We state the assumptions needed regarding the  joint distribution of $(\XX,Y)$. 
 \begin{enumerate}
 	\item[A.1] \label{den_cont} $\bW$ has a density $f_\bW$, which is continuous and positive at $\bw_0$. 
 	\item[A.2]\label{momento_de_Y}
 	There exists $\delta>0$ such that $\esp(\vert Y\vert ^ {2+\delta})<\infty$ and $\esp(\vert Y\vert ^ {2+\delta}\mid \bW=\cdot )$ is continuous at $\bw_0$. 
 	
 	\item[A.3] \label{continuidad} $\regd$, $\regd_2$ and $\sigma^ 2$, defined in (\ref{regd}),  are continuous functions at $\bw_0$ and $\sigma^2(\bw_0)>0$.
 	
 	\item[A.4] \label{ck} There exists $q\geq 2$ such that $f_\bW$ and $\regd$ are of class $C^ q$ on an open convex set containing $\bw_0$. 
 	
 	\item[A.5] \label{reg_extra} $\esp(Y^2\Vert\XX\Vert^2)<\infty$ and, for $a=0,1$,  
 	$\esp \left(Y^a  
 	\XX \mid \bW =\cdot \right)$,  $\esp \left(Y^{2a}  
 	\Vert\XX\Vert^2 \mid  \bW =\cdot \right)$ and $\esp \left(Y^{2a}  
 	\Vert \xx_0-\XX \Vert^ 2 \mid  {\mathbf W} =  \cdot \right)$
	 are continuous at $\bw_0$.   	
 \end{enumerate}
 Now, we present the conditions to be satisfied by  the kernel. 
  \begin{enumerate}
 	\item[K.1] \label{acotado}The kernel  $K:\mathbb R^ d \to \mathbb R$ is bounded,   integrable  with  $\displaystyle{\lim_{\Vert \bw\Vert \to \infty }\Vert \bw K(\bw)\Vert=0}$, and
 	\begin{equation}
 		\label{integra1}  \displaystyle{\int_{\mathbb R^ d} K(\bw)d\bw=1}.
 	\end{equation}

 	\item[K.2]  $
 		\int_{\mathcal R^d}\bv^{\balfa} K(\bv) \;d\bv=0\;, \hbox{for all ${\balfa}$ with $1\leq \vert{\balfa}\vert\leq q-1 \!\!$}, 
$
 for $q$ considered in A.4.
 	
 	\item[K.3]  \label{kequivariante} $K(\bu)=k(\Vert\bu\Vert)$, for $k:\mathbb R\to \mathbb R$ with two continuous derivative and  compact  support satisfying
 	$$
 	\displaystyle{\int_{\mathbb R^ d} k' \left(\left\Vert \bw \right\Vert\right) \frac{\bw^ T}
 		{\left\Vert \bw\right\Vert}\;d\bw=0.}$$
 	
 	\item[K.4] \label{kprima} There exists a positive constant $C>0$ such that  $\vert k' (t) \vert \le C |t| $ {for all $t\in \mathbb R$. }

 \end{enumerate}

%Finally, the relation between the sample size $n$ and the bandith $h$. 
% \begin{enumerate}
% 	\item[R.1] \label{rate_consistencia} The usual condition to get consistency of Nadaraya- Watson estimator when the predictors lie y $\mathbb R^ d$: $h_n\to 0$ and  $nh^d_n\to \infty$ as $n\to \infty$. For $d=1$, we also assume  that $nh^2\to \infty$.
% 	
% 	\item[R.2] \label{rate_0} $nh^{d+2k}\to 0$ as $n\to \infty$.
% \end{enumerate}

An example of such a kernel for $q=2$ is given by $K(\xx) = k(||\xx||) $ with $k(s)=c (1-s^2)^3 I_{(-1<s<1)}$ where $c$ is chosen so that \eqref{integra1} holds.

Condition K.1 includes requirement  \eqref{integra1}, which  is crucial for non parametric density estimation. However, in the regression setting,  the normalization factor required to satisfy \eqref{integra1} cancels both expressions \eqref{est-lim1} and \eqref{est-lim}. 
Thus, in practice, the normalization condition is not needed 
but it makes proofs more elegant.
%but we decided to include it in  lieu of \textcolor{red}{ having carrying} it along the proofs. 
Condition K.3 implies K.1, but we decided to include both of them to emphasize at which point of the proofs each one is used. K.1 and K.2 are typically assumed in non parametric settings. 
%see for  instance \textcolor{red}{\cite{parzen1962estimation}, \cite{nadaraya1964estimating}, \cite{boente1989robust}}. 
 K.3 and K.4 are needed to control the effect of substituting $\bbe_0$ with $\widehat\bbe_n$.  
  \begin{theorem}
  	\label{asymptotic}
  	Assume that $\sqrt{n}(\widehat \bbe_n-\bbe_0)=O_P(1)$. Under  A.1-A.5 and  K.1-K.4   the estimator $\widehat{\eta}(\xx_0)$, defined in  (\ref{est-lim}), consistently  estimates  $\eta(\xx_0)$ if  $h_n\to 0$ and  $nh^d_n\to \infty$ as $n\to \infty$; i.e.
\begin{equation}
	\label{consistencia}
	\widehat{\eta}(\xx_0)\convprob  \eta(\xx_0).
\end{equation}
%if  $h_n\to 0$ and  $nh^d_n\to \infty$ as $n\to \infty$
 For $d=1$, we also assume  that $nh^2\to \infty$. 
Additionally, if $nh_n^{d+2q}\to 0$ as $n\to \infty$, we get that 
  	\begin{equation}
  		\label{dist_asint}
  		\sqrt{nh^d_n}\left\{\widehat\eta(\xx_0)-\eta(\xx_0)\right\}\convdist \mathcal N\left(0, \frac{\sigma^2(\bw_0)  \int_{\mathbb R^d} K^2(\bu)\;d\bu}{f_\bW (\bw_0)}\right)\;.
  	\end{equation}
  	
  \end{theorem}
The relation between $n$ and $h_n$  in Theorem \ref{asymptotic} is that which is typically assumed in the non parametric regression framework when the predictors are in $\mathbb R^ d$. Moreover, in this work we prove that non parametric post dimension reduction procedures based on the estimated reduction behave as well as those based on the  true reduction, in the sense that are asymptotically equivalent. This is a consequence of the different rates of convergence of each procedure. Namely, the $\sqrt{nh^d}$ rate of the non parametric technique kills the impact of replacing the true reduction by its estimator. I.e., it has a better rate of convergence, which is $\sqrt{n}$ in the present case. This is a drastic difference from the parametric setting presented in \cite{kim2020post}, where the use of the estimated sufficient predictors as the true predictors for parametric inference impacts the asymptotic variance of the estimator. For instance, the results presented in the Appendix (Theorem \ref{EL_teo} and Proposition \ref	{asymptotic})  show that we can also estimate non-parametrically the density of  $\bbe_0\XX\mid Y=a$, for $a=0,1$,  based on $(\widehat\bbe_n\XX_i, Y_i)$, with the same rates as if the reduction were known. This can be used in classification tasks by invoking a generative approach, i.e. when the conditional distribution  of $X \mid Y$ is used instead of $Y \mid X$.

\subsection{Non linear regression model}

One of the examples where this theory can be applied is when $\esp(Y\mid\XX)$ is a non linear function 
of some linear combinations of $\XX$. This  is a particular case of the  non linear regression model, which postulates that 
\begin{equation}
	\label{non_linear}
	Y=\gnl(\bbe_0\XX)+\varepsilon, 
\end{equation}
where $\bbe_0\in \mathbb R^{d\times p}$, $\XX\in \mathbb R^p$, $\gnl:\mathbb R^ d\to \mathbb R$ is an unknown function, 
$\XX$  is independent of $\varepsilon$, $\esp(\varepsilon)=0$ and $\esp(\varepsilon^ 2)=\sigma^2$. 
This model fits the framework proposed in \eqref{lin_red}. In this case,  following the definitions presented in \eqref{regd}, we get that
\begin{equation}
	\regd(\bw)=\gnl(\bw)\;, \;\;	\regd_2(\bw)=\gnl^2(\bw)\;,\;\;\sigma^2(\bw)=\sigma^ 2\;.
\end{equation}

The  assumptions A.1-A.5 can now  be restated as follows.

\begin{enumerate}
	\item[A.1] \label{den_cont} $\bW$ has a density $f_\bW$, which is continuous and positive at $\bw_0$. 
	\item[A.2]\label{momento_de_Y}
	There exists $\delta>0$ such that $\esp(\vert \gnl(\bbe_0\XX)\vert ^ {2+\delta})<\infty$ and $\gnl$  is continuous at $\bw_0$. 
	
	\item[A.3] \label{continuidad} $\gnl$ is continuous  at $\bw_0$ and $\sigma^2>0$.
	
	\item[A.4] \label{ck} There exists $q\geq 2$ such that $f_\bW$ and $\gnl$ are of class $C^ q$ on an open convex set containing $\bw_0$. 
	
	\item[A.5] \label{reg_extra} $\esp(\Vert \XX\Vert ^2)<\infty$, $\esp(\gnl^2(\bbe_0\XX)\Vert\XX\Vert^2)<\infty$. The functions 
	$\esp \left(\XX\mid \mathbf{W}=\cdot \right)$,  $\esp \left(\Vert \XX\Vert ^2\mid \mathbf{W} =\cdot \right)$
	and $\gnl$  are continuous at $\bw_0$.

\end{enumerate}

%\textcolor{red}{donde hay resultados de consistencia para estimar $\bbe_0$?}

 \subsection{Uniform Consistency}\label{consistencia}
 
 Theorem \ref{asymptotic} guarantees the consistency of the non-parametric estimation under some conditions.
 %As in \eqref{A_n_sombrero}, from Theorem \ref{EL_teo} and Lema \ref{ktaylor}, we get that 
% $$
% \left\vert \frac{1}{nh_n^d}\sum_{i=1}^ nY_i^a K\left(\frac{\widehat \bbe_n (\xx_0-\XX)}{h_n}\right)-\regd^a(\bbe_0\xx_0)f_\bW(\bbe_0\xx_0)\right\vert\convprob   0.
% $$
 In this section we will show that this convergence is preserved when taking the supreme with $\xx_0$ varying in a compact set $\mathcal K$ if certain conditions are met. The proof is based on  results established in the empirical processes literature. 
%As often happens in non parametric settings, to prove the convergences of each sequence  presented  in (\ref{cada_uno}), 
% we  use analysis arguments  
% to handle  the bias terms (see Appendix),
% while the remaining random components are  considered  in Theorem \ref{desig-max}, invoking   some results established in the empirical processes literature. 
 %
 %
 %For $a\in \{0,1\}$ define 
 %\begin{equation}
 %\label{Sh}
 % S_{a}(\bbe,\xx_0,h)=\frac{1}{h^d_n} \EE\left[Y^aK\left(\frac{\bbe^\trans \xx-\bbe^\trans\xx_0}{h}\right)\right]\;,
 %\end{equation}
 %and consider 
 %\begin{equation}
 %\label{losS}
 %S_{0}(\bbe,\xx_0)=f(\bbe^\trans\xx_0,\bbe)\;,\quad  S_{1}(\bbe,\xx_0)=\EE[Y\vert \bbe^\trans \XX=\bbe^\trans\xx_0]f(\bbe^\trans\xx_0, \bbe).
 %\end{equation}
 We will assume that: 
 \begin{enumerate}

 	\item[S.1] \label{bounded_variation}  $K (m) = k (\Vert m \Vert)$, for $k $ bounded, with compact support of bounded variation.
 	
 	\item[S.2] \label{den_cont_unif} The density function $f(t,\bbe)$ of $\bbe \XX$ is continuous in $(t,\bbe)$ and  for every compact set $\mathcal{K}\subset \mathbb{R}^{p}$
 	\begin{equation}
 		\label{den_unif_bounded}\inf_{\xx \in \mathcal{K}}  f(\bbe_0 \xx,\bbe_0)=
 		\inf_{\xx \in \mathcal{K}}  f_{\bW}(\bbe_0 \xx)>0.
 	\end{equation}
 	\item[S.3] \label{reg_cont_unif}  The regression function $r(t,\bbe)= \EE[Y\vert \bbe \XX=t]$ is continuous 
 	in $(t,\bbe)$.
 	
 \end{enumerate}
 
 Note that S.1 is satisfied, for example, if $K(\mathbf{z})=k(\Vert \mathbf{z}\Vert)$, where $k$ is any of the following: the uniform kernel, the Epanechnikov kernel, the biweight kernel or the triweight kernel.

 \begin{theorem}
 	\label{consistency}
 	Assume that $\widehat\bbe_n \cp \bbe_0$ and  that S.1-S.3 hold.  Let $\mathcal{K} \subset \mathbb{R}^{p}$ be a compact set.
 	If either
	\begin{enumerate}
 		\item There exists a constant $C$ such that $\vert Y \vert \leq C$ with probability one, $h_n \rightarrow 0$ and $h_{n}^{d}n/\log n \rightarrow \infty$.
 		\item $\EE Y^{2} <\infty$, $h_n \rightarrow 0$ and $h_{n}^{2d}n \rightarrow \infty$,
 	\end{enumerate}
	Then $\sup_{\mathbf{x}_0 \in \mathcal{K}} \vert \widehat\eta(\xx_0)-  \eta(\xx_0) \vert \cp 0$.

 \end{theorem}

% \section{\textcolor{red}{Other Functionals}}
 %\textcolor{red}{aca se puede hablar con Graciela. yo creo que seria lindo sumar algo de la mediana. quede trabada con las cuentas cuando aparece el supremo. no supe avanzar
% Let $F(y\mid \XX=\xx)$ denotes the conditional distribution of $Y\mid \XX=\xx$. 
 % \begin{proposition} Under the same conditions of Theorem \ref{asymptotic} 
% 	\label{boundedktaylor}
% 	\begin{equation}
% 		\label{each}
% 		\sqrt{nh^d_n}\sup_{y\in \mathbb R^ d}\left\vert\frac{1}{nh^ d}	\sum_{i=1}^n  I^a_{Y_i\leq y} \left\{ K\left(\frac{ \bbe_0\xx_0-\bbe_0\XX_i}{h}\right)-
 %		K\left(\frac{ \widehat \bbe_n\xx_0-\widehat\bbe_n\XX_i}{h}\right)\right\}\right\vert=o_P(1)\;,\quad\hbox{for $a=0,1$}\;.
% 	\end{equation}
% \end{proposition}
%}

 \section{Projections} \label{proy-1}
 In the literature of SDR, given $(\XX,Y)\in \mathbb R^{p+1}$ there are methods to characterize 
 an orthogonal projection of rank $d<p$,  $\proy_0\in \mathbb R^{p\times p}$, such that 
 \begin{equation}
 	\label{lin_red_2}
 	 Y\mid \XX\sim Y\mid \proy_0\XX.
 \end{equation}
 See for example \cite{li2003dimension, cook_forzani_2008,cook2021pls,li91,cookweisberg,li92,bura01,Xia02,Li05,CookNi,Zhu06,CookLi,LiWang}.
  In particular, (\ref{lin_red_2}) implies
 \begin{equation}\label{ultima}
 \esp ( Y \mid \XX ) = \esp (Y \mid\proy_0 \XX).
 \end{equation}
 Therefore for any fixed $\bbe_0\in \mathbb R^{d\times p}$,  
 with 
 $\proy_0\bbe_0^T=\bbe_0^T$ and $\bbe_0\bbe_0^T=I_d$, 
  $$\esp ( Y \mid \XX ) = \esp (Y \mid\proy_0 \XX)= \esp (Y\mid \bbe_0 \XX),
  $$
 and therefore, we fit the setting presented in Section \ref{est_reg_red}. 
 %or 
 %$Y\mid \XX\sim Y\mid \proy_0\XX$ that implies \eqref{ultima}  
 % is characterized  
 
 Asymptotically normal estimators $\widehat \proy_n$ of $\proy_0$, satisfying  (\ref{lin_red_2}), are presented in the same literature.  However, nothing is mentioned   about the existence of $\widehat {\bbe}_n$ converging to $\bbe_0$, as we assumed in the construction of $\widehat \eta(\bx_0)$, presented in (\ref{est-lim}).
 Fortunately, when using an invariant kernel  $K$ as in S.1 from Section \ref{consistencia} or K.3  we have $K(A\xx)=K(\xx)$ for any $A\in \mathbb R^{d\times d}$ with $AA^T=I_d$, and 
  the definition of $\widehat \eta(\xx_0)$ presented in display (\ref{est-lim}) does not depend on $\widehat{\bbe}_n$ but on  $\spn (\widehat{\bbe}_n^T)$. Therefore,  we only need the existence of one of such sequence $\widehat \bbe_n$ and then any orthonormal matrix with the same span as $\widehat \proy_n$ can be used to  construct $\widehat \eta(\bx_0)$. The existence of such a sequence is guaranteed by the following proposition.

 \begin{proposition}
 	\label{sequence}
 
 	Assume that $\widehat \proy_n\in \mathbb R^{p\times p}$,  and $\sqrt{n}(\widehat\proy_n-\proy_0)  =O_P(1)$, with $\hbox{rank}(\widehat \proy_n)=d$,  $\hbox{rank}(\proy_0)=d$. 
 Fix $\bbe_0$
 such that
 $\proy_0\bbe_0^T=\bbe_0^T$ and $\bbe_0\bbe_0^T=I_d$. Then, there exists $\widehat \bbe_n$ such that  	
 	$\widehat \bbe_n^T$ is a basis of $span(\widehat \proy_n)$: $\widehat {\bbe}_n\widehat {\bbe}_n^T=I_{d}$, $\widehat\proy_n\widehat {\bbe}_n^T=\widehat {\bbe}_n^T$ and $\sqrt{n}(\widehat{\bbe}_n-\bbe_0)=O_P(1).$ 
 \end{proposition}
 As a consequence, if a sufficient dimension reduction estimator is found for the regression of $Y \mid X$ we could use 
 (\ref{est-lim}) to predict $Y$ for a new value $\xx_0$ of the predictors, and we get an efficient estimator as if 
 the true reduction were known.

 \section{Sufficient dimension reduction}
 
 \subsection{Monte Carlo Study}\label{simu}
 
 %SimulacionModelo1VariandoN
 
 We report the results of a Monte Carlo simulation study 
 where we compute $\widehat {\mathbb E}(Y\mid \mathbf X=\bx)$, in three different ways: (i) performing a kernel regression on the original space of covariates, from now on \texttt{NP} (ii) with a kernel regression based on the true $\bbe_0$, as indicated in (\ref{est-lim1}), from now on \texttt{NPR} and (iii) with the recipe presented in (\ref{est-lim}), where $\widehat{\bbe}_n$ is a $\sqrt{n}$ consistent estimator of $\bbe_0$, from now on \texttt{NPRT}. Since in practice $\bbe_0$ is unknown, the second procedure is unfeasible.

%   with 
 
 	\subsubsection{Model 1. Forward regression} 
In this scenario,  the predictors are generated as 
$\mathbf{X}\sim\mathcal{N}(0, \Sigma)\in \mathbb R^p$ with $p=6$, $\Sigma = 5\bbe_0 \bbe_0^T + 0.1(I_6 - \bbe_0  \bbe_0^T)$ with $\bbe_0 = 1_6/ \sqrt{6}$,  while the response is given by 
$Y=(\bbe_0^T  \mathbf{X})^2 + \epsilon$ with $\epsilon \sim N(0, (0.5)^2)$ and therefore the sufficient reduction is given by $\bbe_0^T \mathbf{X}$.

\begin{figure}[ht!]
 	\centering
 	\includegraphics[width=2.3in,height=1.6in]{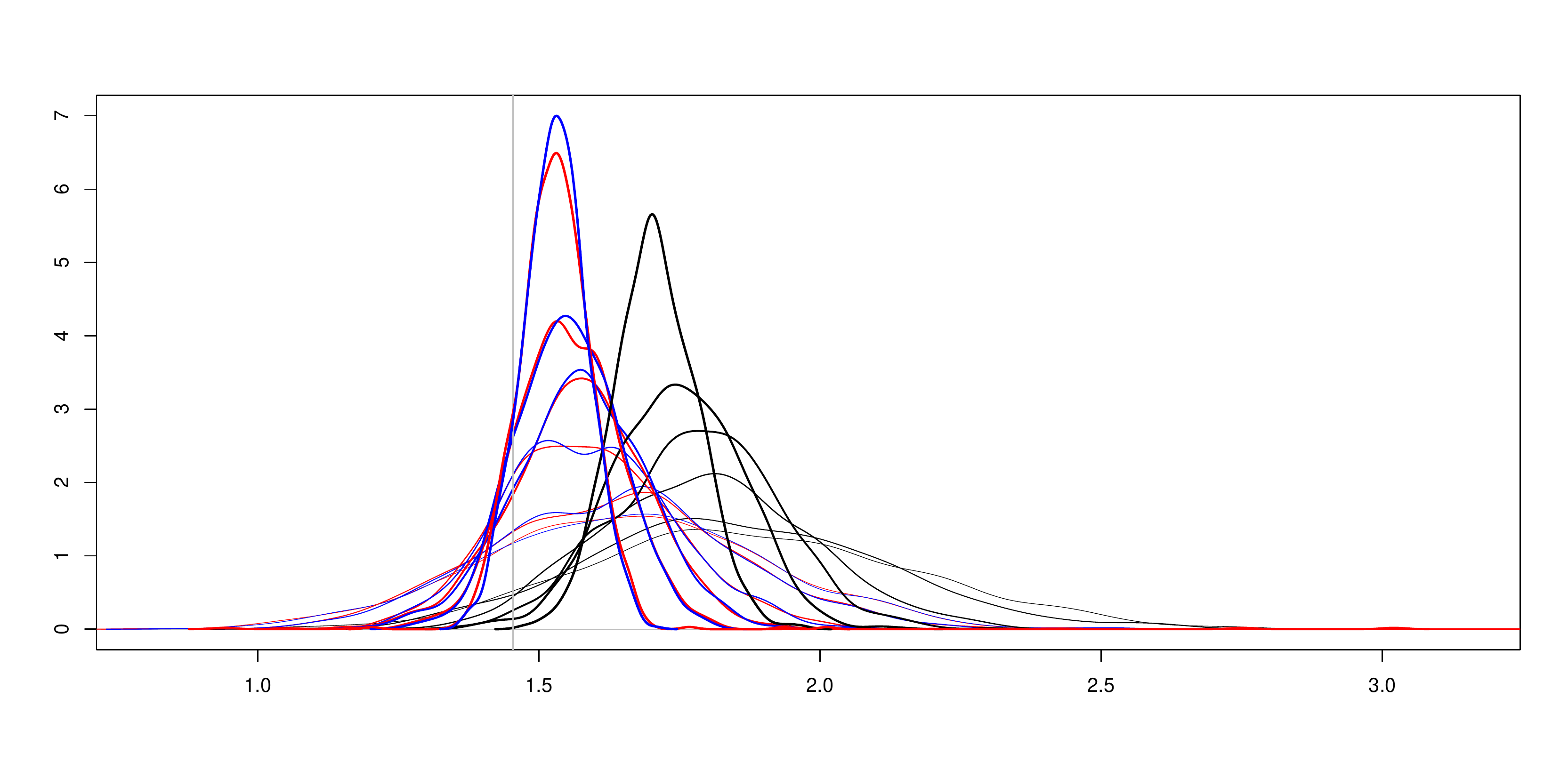}\includegraphics[width=2.3in,height=1.6in]{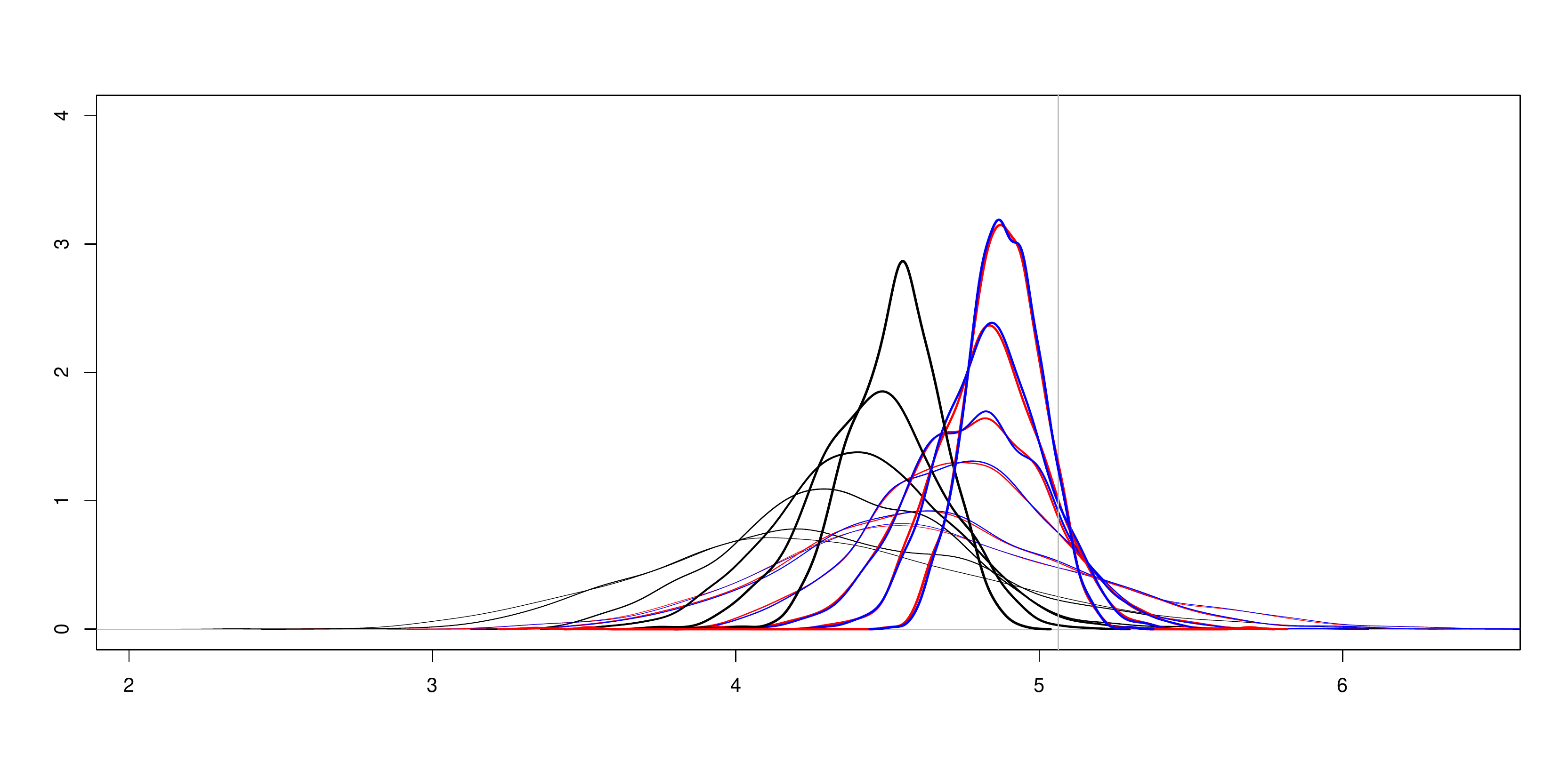}
 	\caption{Density estimation of the non-parametric after reduction  (red),   density estimation of  non-parametric regression without reduction (black), and  density estimation of  non-parametric regression using the true  reduction (blue). All the estimations were computed  at the same two random points (each figure is one of the two points).}
 	\label{fig:modelo1punto1y9}
 \end{figure}

 We generate samples of size $n= 80, 100, 200, 300, 500,  1000$. For each of them we estimate the regression function $\mathbb E(Y\mid \mathbf X=\bx)$   with 	$\widehat {\mathbb E}(Y\mid \mathbf X=\bx)$, computed in the three  different ways where for \texttt{NPR}, $\widehat{\bbe}_n$ is the first PLS component, computed with \cite{varmuza} that we know is sufficient using \cite{cook2021pls}.  In each case, the bandwidth $h$ has been chosen as $5 n^{-1/(4+p)}$ and the kernel $k$ is given by 
\begin{equation}
	\label{kernel}
k(x)=(1-x^ 2)^3 /{\mathcal{B}}(0.5,4)\;I_{-1\leq x\leq 1},
\end{equation}
where $\mathcal{B} $ is the Beta function.
We compare the mean square error of the estimators   $\widehat{\mathbb E}(Y\mid \mathbf X=\bx)$  with $  {\mathbb E}(Y\mid \mathbf X=\bx)$
using the empirical mean square error
given by 
\begin{equation}
	\hbox{EMSE}(\bx, n)=\frac{1}{N_{rep}}\sum_{i=1}^{N_{rep}}\left\{\widehat {\mathbb E}(Y\mid \mathbf X=\bx) -\frac{1}{N_{rep}}\sum_{i=1}^{N_{rep}}\widehat {\mathbb E}(Y\mid \mathbf X=\bx)\right\}^2,
\end{equation}
where $N_{rep}=1000$ stands for the number of replications performed. 
We consider 10 fixed points, $\bx=\bx_{test,j} \in {\mathbb R}^p, j=1,\dots,10$, randomly chosen following the distribution of ${\mathbf X}$.
Table \ref{lab-ecm} exhibits the $\hbox{EMSE}(\bx_{test,j}, n)$, $j=1,\dots,10$, $n= 80, 100, 200, 300, 500,  1000$ where,  as expected, the  
$\hbox{EMSE}$ decreases with  the sample size and also if we reduce  predictors before performing the non parametric estimator. It also shows that there is not loss of information if we consider the estimated $\widehat{\bbe}_n$ instead of $\bbe_0$.
We also include in Figure \ref{fig:modelo1punto1y9} the density plots of the estimates of the regression function at   $\bx_{test,j}$, $j=1$ and $j=9$ where the distribution is normal and again, as expected the variance decreases drastically when reduction is considered.

\begin{table}[!ht]
\centering
\begin{tabular}{rlrrrrrr}
  \hline\hline
&Method &$n=80$ & $n=100$ & $n=200$ & $n=300$ & $n=500$ &$n=1000$ \\ 
  \hline\hline
 & \texttt{NP} & 0.2688 & 0.2300 & 0.1507 & 0.1274 & 0.0974 & 0.0695 \\ 
$\bx_{test,1}$&\texttt{NPR} & 0.1029 & 0.0861 & 0.0598 & 0.0464 & 0.0224 & 0.0097 \\ 
 &\texttt{NPRT} & 0.1028 & 0.0801 & 0.0389 & 0.0282 & 0.0175 & 0.0089 \\ \hline
& \texttt{NP} & 0.1510 & 0.1120 & 0.0565 & 0.0371 & 0.0230 & 0.0103 \\ 
$\bx_{test,2}$&\texttt{NPR}     & 0.1090 & 0.0835 & 0.0490 & 0.0353 & 0.0204 & 0.0113 \\ 
 &\texttt{NPRT}  & 0.1061 & 0.0803 & 0.0361 & 0.0248 & 0.0143 & 0.0056 \\ \hline
  & \texttt{NP} & 1.2488 & 1.1169 & 0.8314 & 0.6659 & 0.5361 & 0.4165 \\ 
$\bx_{test,3}$&\texttt{NPR}    & 0.5517 & 0.4639 & 0.2407 & 0.1586 & 0.1002 & 0.0571 \\ 
   &\texttt{NPRT}& 0.5352 & 0.4481 & 0.2359 & 0.1526 & 0.0955 & 0.0554 \\ \hline
  & \texttt{NP} & 0.4090 & 0.3756 & 0.2713 & 0.2240 & 0.1789 & 0.1322 \\ 
  $\bx_{test,4}$&\texttt{NPR}  & 0.1220 & 0.1184 & 0.0854 & 0.0702 & 0.0296 & 0.0158 \\ 
   &\texttt{NPRT}& 0.1225 & 0.1050 & 0.0566 & 0.0382 & 0.0249 & 0.0143 \\ \hline
  & \texttt{NP} & 0.2382 & 0.2019 & 0.1281 & 0.1064 & 0.0801 & 0.0561 \\ 
 $\bx_{test,5}$&\texttt{NPR}   & 0.1017 & 0.0864 & 0.0587 & 0.0499 & 0.0336 & 0.0094 \\ 
   &\texttt{NPRT}& 0.1001 & 0.0773 & 0.0370 & 0.0268 & 0.0164 & 0.0081 \\ \hline
  & \texttt{NP}& 0.4215 & 0.3863 & 0.2801 & 0.2311 & 0.1845 & 0.1370 \\ 
 $\bx_{test,6}$&\texttt{NPR}   & 0.1236 & 0.1222 & 0.0853 & 0.0794 & 0.0377 & 0.0156 \\ 
   &\texttt{NPRT}& 0.1242 & 0.1064 & 0.0575 & 0.0390 & 0.0254 & 0.0147 \\ \hline
  & \texttt{NP} & 13.5624 & 12.7395 & 9.5869 & 8.3357 & 6.6559 & 4.9766 \\ 
 $\bx_{test,7}$&\texttt{NPR}   & 5.1656 & 4.6299 & 2.4306 & 2.0336 & 1.2040 & 0.6107 \\ 
   &\texttt{NPRT}& 5.0465 & 4.3090 & 2.2778 & 1.7404 & 1.0990 & 0.5945 \\ \hline
  & \texttt{NP} & 0.8925 & 0.8295 & 0.6284 & 0.5351 & 0.4365 & 0.3340 \\ 
$\bx_{test,8}$&\texttt{NPR}    & 0.2223 & 0.2025 & 0.1537 & 0.1233 & 0.0617 & 0.0318 \\ 
   &\texttt{NPRT}& 0.2207 & 0.1858 & 0.1062 & 0.0773 & 0.0512 & 0.0303 \\ \hline
  & \texttt{NP} & 0.9691 & 0.8685 & 0.6346 & 0.4972 & 0.3997 & 0.3060 \\ 
 $\bx_{test,9}$&\texttt{NPR}   & 0.4539 & 0.3834 & 0.1970 & 0.1300 & 0.0836 & 0.0477 \\ 
   &\texttt{NPRT}& 0.4412 & 0.3704 & 0.1923 & 0.1219 & 0.0764 & 0.0441 \\ \hline
  & \texttt{NP} & 0.7523 & 0.6900 & 0.5142 & 0.4414 & 0.3580 & 0.2719 \\ 
  $\bx_{test,10}$&\texttt{NPR}  & 0.1928 & 0.1785 & 0.1326 & 0.1068 & 0.0546 & 0.0264 \\ 
  &\texttt{NPRT} & 0.1912 & 0.1616 & 0.0898 & 0.0652 & 0.0432 & 0.0255 \\ \hline\hline
\end{tabular}
\caption{Mean square error for the nonparametric estimation computed at 10 random points, acording to Model 1.}\label{lab-ecm}
\end{table}

% \subsubsection{Modelo 3}
%
%Copio los resultados en bruto para ver si vale la pena incluirlos
%lo que veo es que los resultados tienen mucho sesgo.
%la Y nueva es normal con media 0 y varianza 5, los resultados que dan bien corresponden a valores grandes de Y positivos, a medida que el y se hace cada vez mas chico empieza a ganar sesgo. En los graficos se ve mejor
%
%esto es el no parametrico
%
\begin{table}[!ht]
\centering
\begin{tabular}{rlrrrrrr}
  \hline\hline
&Method &n=80 & n=100 & n=200 & n=300 & n=500 &n=1000 \\ 
  \hline\hline
&\texttt{NP}  & 0.9683 & 0.8810 & 0.7213 & 0.5609 & 0.5329 & 0.4008 \\ 
$\bx_{test,1}$& \texttt{NPR} & 0.7606 & 0.5861 & 0.3275 & 0.3310 & 0.3006 & 0.2746 \\ 
&\texttt{NPRT} & 0.1351 & 0.1209 & 0.0914 & 0.0821 & 0.0767 & 0.0627 \\ \hline
  &\texttt{NP} & 2.0874 & 1.7769 & 1.0336 & 1.0175 & 0.7051 & 0.5274 \\ 
  $\bx_{test,2}$& \texttt{NPR} & 0.9703 & 0.6480 & 0.3757 & 0.4076 & 0.3092 & 0.2991 \\ 
  &\texttt{NPRT} & 0.5817 & 0.5420 & 0.3218 & 0.2931 & 0.3016 & 0.2409 \\ \hline
  &\texttt{NP} & 2.3966 & 1.8517 & 1.6623 & 1.4669 & 1.2223 & 0.8977 \\ 
  $\bx_{test,3}$& \texttt{NPR} & 1.4915 & 1.0441 & 0.5305 & 0.4645 & 0.4011 & 0.3096 \\ 
  &\texttt{NPRT} & 0.0929 & 0.0912 & 0.0705 & 0.0664 & 0.0675 & 0.0608 \\ \hline
  %cambie el 4 por el 11
  %&\texttt{NP} & 3.6229 & 3.1008 & 2.5219 & 2.4615 & 2.4721 & 1.9656 \\ 
  %$\bx_{test,4}$& \texttt{NPR} & 3.3275 & 3.1301 & 2.7598 & 2.6544 & 2.5422 & 2.3591 \\ 
  %&\texttt{NPRT} & 0.0870 & 0.0929 & 0.0686 & 0.0682 & 0.0672 & 0.0605 \\ \hline
  &\texttt{NP} & 2.0874 & 1.7769 & 1.0336 & 1.0175 & 0.7051 & 0.5274 \\ 
  $\bx_{test,4}$& \texttt{NPR} & 0.9703 & 0.6480 & 0.3757 & 0.4076 & 0.3092 & 0.2991 \\ 
  &\texttt{NPRT} & 0.5817 & 0.5420 & 0.3218 & 0.2931 & 0.3016 & 0.2409 \\ \hline
  &\texttt{NP} & 2.0874 & 1.7769 & 1.0336 & 1.0175 & 0.7051 & 0.5274 \\ 
  $\bx_{test,5}$& \texttt{NPR} & 0.9703 & 0.6480 & 0.3757 & 0.4076 & 0.3092 & 0.2991 \\ 
  &\texttt{NPRT} & 0.5817 & 0.5420 & 0.3218 & 0.2931 & 0.3016 & 0.2409 \\ \hline
  &\texttt{NP} & 1.5921 & 1.5265 & 1.1053 & 1.0146 & 0.8493 & 0.6286 \\ 
  $\bx_{test,6}$& \texttt{NPR} & 1.0549 & 0.7626 & 0.4268 & 0.3783 & 0.3279 & 0.2862 \\ 
  &\texttt{NPRT} & 0.1687 & 0.1614 & 0.1172 & 0.1214 & 0.1000 & 0.0764 \\ \hline
  &\texttt{NP} & 1.5416 & 1.4425 & 0.9471 & 0.9072 & 0.7703 & 0.5313 \\ 
  $\bx_{test,7}$& \texttt{NPR} & 0.9702 & 0.7283 & 0.4056 & 0.3687 & 0.3302 & 0.2830 \\ \hline
  &\texttt{NPRT} & 0.1706 & 0.1803 & 0.1366 & 0.1147 & 0.0951 & 0.0710 \\
  %cambien el 13 por el 7
  %&\texttt{NP} & 2.7692 & 2.4779 & 1.6848 & 1.6003 & 1.4909 & 1.1712 \\ 
  %$\bx_{test,7}$& \texttt{NPR} & 2.8413 & 2.4547 & 1.9807 & 2.2614 & 2.1281 & 2.2976 \\ 
  %&\texttt{NPRT} & 1.5474 & 1.4265 & 1.0913 & 0.9541 & 0.9648 & 0.7189 \\ \hline
  &\texttt{NP} & 2.0874 & 1.7769 & 1.0336 & 1.0175 & 0.7051 & 0.5274 \\ 
  $\bx_{test,8}$& \texttt{NPR} & 0.9703 & 0.6480 & 0.3757 & 0.4076 & 0.3092 & 0.2991 \\ 
  &\texttt{NPRT} & 0.5817 & 0.5420 & 0.3218 & 0.2931 & 0.3016 & 0.2409 \\ \hline
  &\texttt{NP} & 3.6533 & 2.8136 & 2.5971 & 2.3095 & 2.0781 & 1.6019 \\ 
  $\bx_{test,9}$& \texttt{NPR} & 1.9848 & 1.5497 & 1.0104 & 0.8021 & 0.7012 & 0.4987 \\ 
  &\texttt{NPRT} & 0.0909 & 0.0902 & 0.0675 & 0.0681 & 0.0684 & 0.0609 \\ \hline
  &\texttt{NP} & 1.5576 & 1.1734 & 1.0350 & 0.8164 & 0.6965 & 0.5073 \\ 
  $\bx_{test,10}$& \texttt{NPR} & 1.2073 & 0.7974 & 0.3682 & 0.3606 & 0.3259 & 0.2763 \\ 
  &\texttt{NPRT} & 0.1014 & 0.0923 & 0.0710 & 0.0694 & 0.0698 & 0.0591 \\ 
  %&\texttt{NP} & 2.7695 & 2.4781 & 1.6851 & 1.6007 & 1.4913 & 1.1715 \\ 
  %$\bx_{test,12}$& \texttt{NPR} & 2.8418 & 2.4554 & 1.9813 & 2.2620 & 2.1287 & 2.2982 \\ 
  %&\texttt{NPRT} & 1.5456 & 1.4250 & 1.0898 & 0.9527 & 0.9635 & 0.7177 \\\hline
  \hline\hline
\end{tabular}
\caption{Variance for the nonparametric estimation computed at 10 random points, acording to Model 2.}\label{lab-ecm2}
\end{table}

 \begin{figure}[ht!]
    \centering
\includegraphics[width=2.3in,height=1.6in]{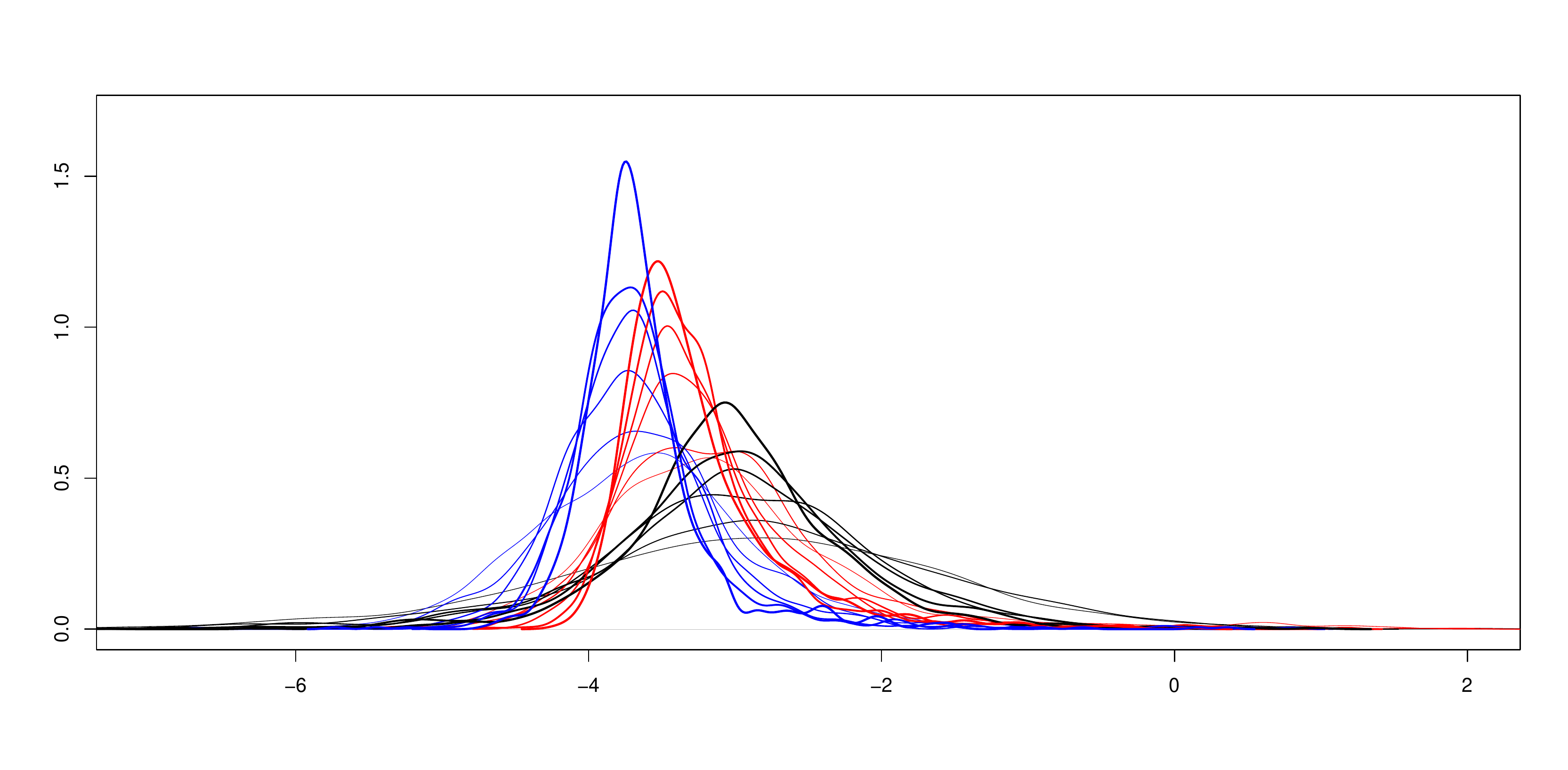}\includegraphics[width=2.3in,height=1.6in]{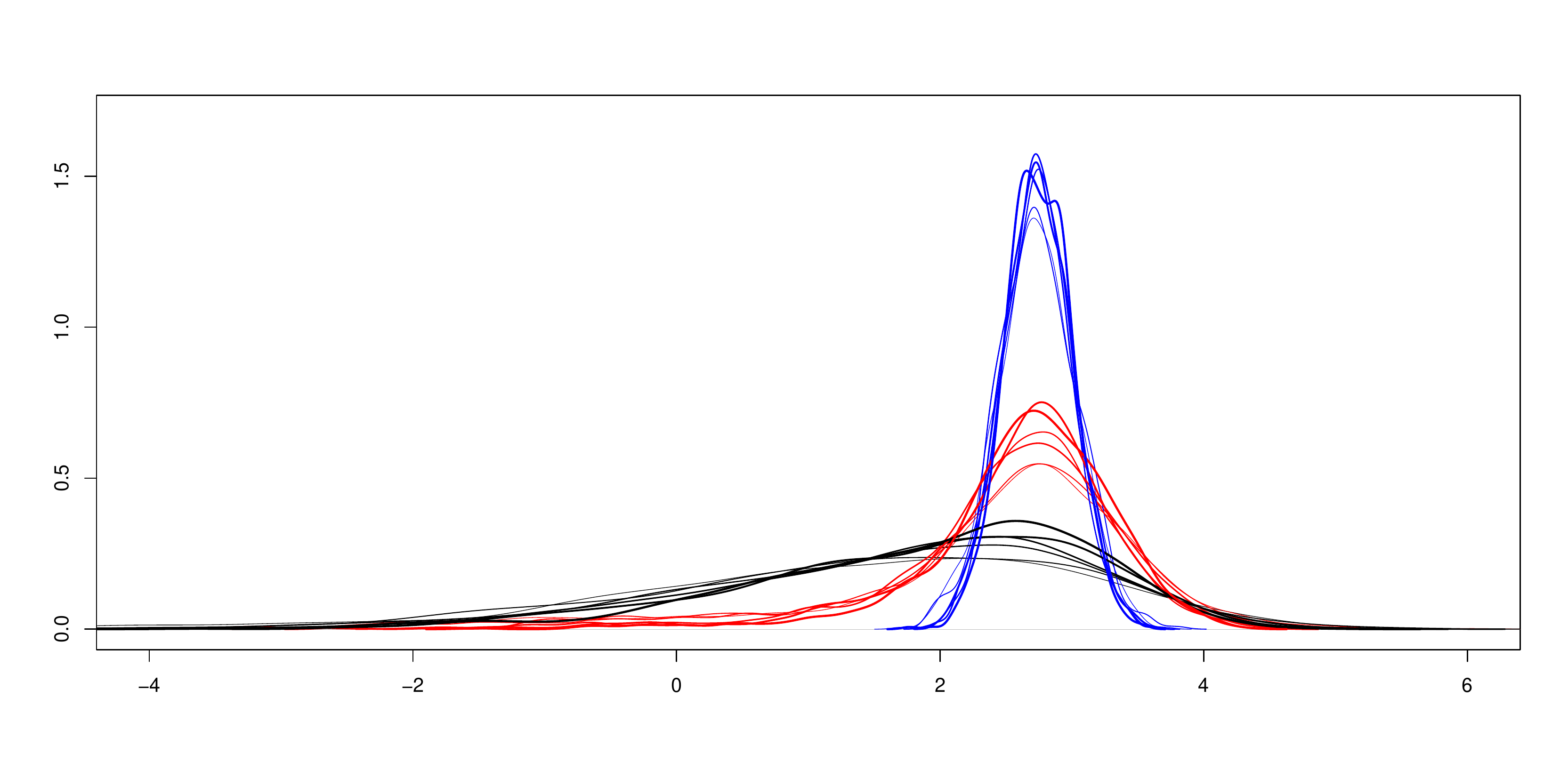}\\\
\caption{Density estimation of the non-parametric after reduction  (red),   density estimation of  non-parametric regression without reduction (black), and  density estimation of  non-parametric regression using the true  reduction (blue). All the estimations were computed  at the same two random points (each figure is one of the two points).}% MODELOS 4  }
\label{fig:modelo2punto1y9}
\end{figure}

\subsection{Model 2. Inverse regression}
In this case, we generate data under an inverse regression model, which postulates that $Y\sim \mathcal N(0, 5^2)\in \mathbb R$ while  the conditional distribution of $\mathbb X\in \mathbb R^{20}$ given $Y=y$ is Gaussian with mean $\nu_y=ABf_y$,  where $f_y=(y-E(Y),\vert y\vert-\esp(\vert Y\vert))^t\in \mathbb R^2$, $A= (1,1,-1,-1,\underbrace{0,\ldots,0}_{\text{16}})/2$, $B=(1,1)^T$ and $\Delta = 0.1 S S^T $ where $S$ was generated onset as a square matrix of dimension 20 of standard normal.
%
%y las matrices $A$ y $B$ estan dadas por  bla bla bla. Namely, $\mathbf{X}\mid Y=y\sim \mathcal N(\nu_y, \Delta)\in \mathbb R^{20}$.
In this case, \cite{cook_forzani_2008} proved that the reduction $\bbe_0$ is given by $\Delta^{-1} A$ and also showed that it can be estimated conssistenly, with $\sqrt{n}$ rate, using principal fitted components.

In each case, the bandwidth $h$ has been chosen as $10 n^{-1/(4+p)}$ and the kernel $k$ is the one used in Model 1.
%\begin{equation}
%	\label{kernel}
%k(x)=(1-x^ 2)^3 /{\mathcal{B}}(0.5,4)\;I_{-1\leq x\leq 1},
%\end{equation}
%where $\mathcal{B} $ is the Beta function.
Table \ref{lab-ecm} exhibits the  variance of the estimators   $\widehat{\mathbb E}(Y\mid \mathbf X=\bx)$  in 10 fixed points, $\bx=\bx_{test,j} \in {\mathbb R}^p, j=1,\dots,10$. As expected the variance decreases with the sample size and when the dimension is reduced (here from $p=20$ to $d=1$), and the loss of information using the estimator of the sufficient dimension reduction, even if it is bigger than in Model 1, still is not significative.

We also include in Figure \ref{fig:modelo2punto1y9} the density plots of the estimates of the regression function at   $\bx_{test,j}$, $j=2$ and $j=9$ where again,  the distribution is normal and as expected the variance decreases drastically when reduction is considered.

% on $Y$ follows a reduced rank generalized linear model. In detail, $Y $ is a discrete random variable taking values in $\{1,\dots, r\}$ with $r=6$ and probability $1/r$ for each point. Then $\XX \mid Y=y \sim\mathcal{N}( \mu_y, \Delta)$ with $\Delta= I_p$, $p=30$ and $\mu_y = \beta f_y$ for  $\beta = {\mathbf A} B$, with $\beta\in \mathbb R^{p\times (r-1)}$,  ${\mathbf A}= {\mathbf 1_p}/\sqrt{p} \in \mathbb R^{p\times 1}$ and $B = 5 {\mathbf 1^{\top}_{r-1}}\in \mathbb R^{1\times(r-1)}$;  $f_y \in {\mathbb R}^{r-1}$ and its {$k$th} coordinate is given by  $f_{yk} = \mathbf{1}( y =k) -1/r$ for $ k \in \{ 1,\dots, r-1\}$. In practice, the center term $1/r$ is replaced by its estimator $n_k/n$,  where $n_k$ is the cardinal of the set $\{ y =k\}$ and $n$ is the sample size. 

 \subsection{Mussels' data}\label{ejemplo}
 
\begin{figure}[ht!]
    \centering
        \includegraphics[width=4.8in,height=2.3in]{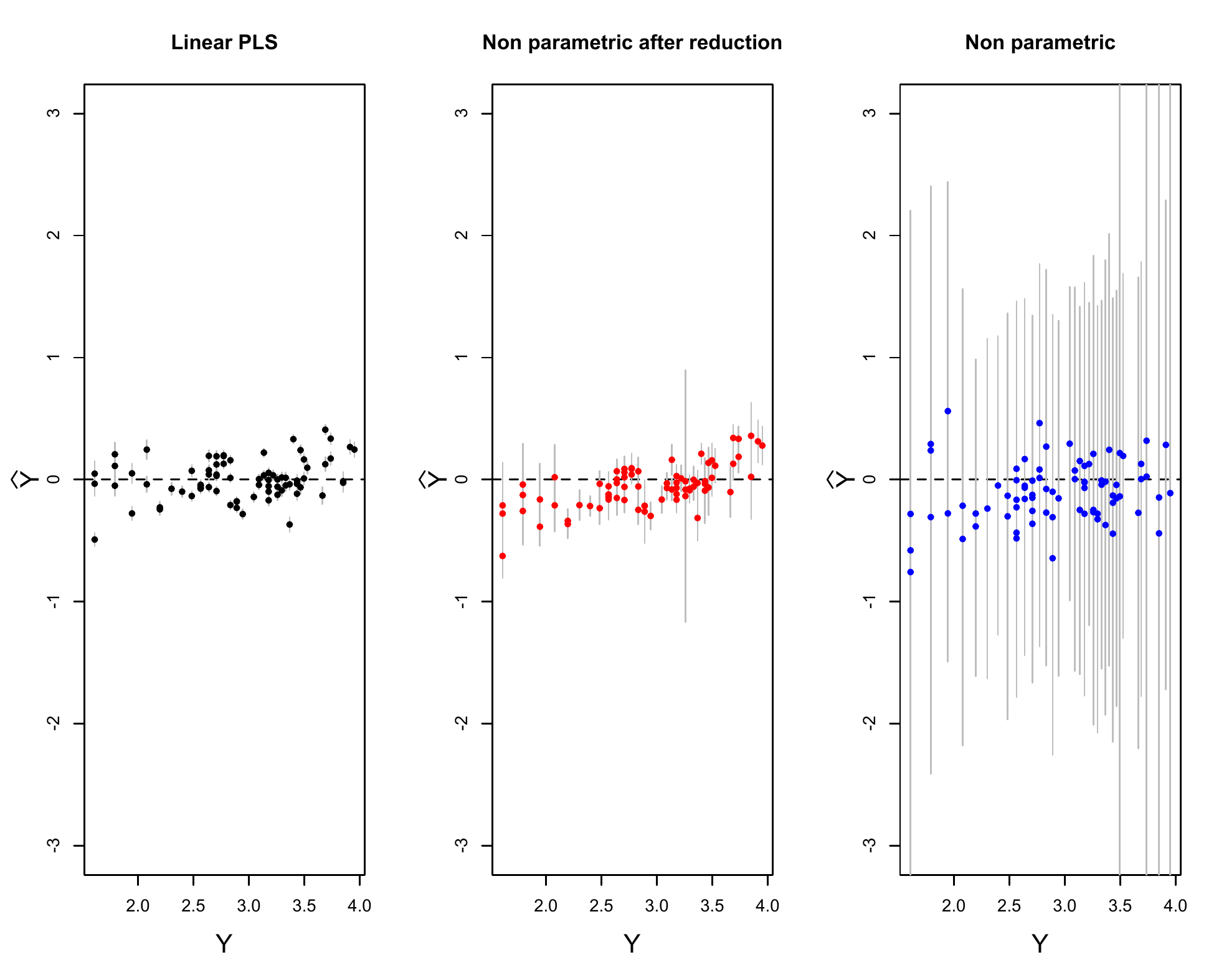}
               \caption{Mussels' data: Plot of the observed responses $Y_{i}$ versus the fitted values $\hat{Y}_{i}$ from the PLS linear fit with one component, $q=1$ (left), non-parametric fit with confidence intervals after reduction (middle) and non-parametric regression without reduction (right)}
        \label{fig:musselsfit}
\end{figure}

 \cite{cook2018introduction}[Section 4.4.4] and \cite{cook2023libro}
 used  Partial Least Square (PLS) linear regression to analyze data from an ecological study of horse mussels sampled from the Marlborough Sounds off the coast of New Zealand. The response variable $Y$ is the logarithm of the mussel's muscle mass, and four predictors ($X$) are the logarithms of  shell height, shell width, shell length (each in millimeters), and shell mass (in grams). The sample size is $n=79$ and $p=4$. \cite{cook2018introduction} obtained a clear inference that the regression required only a single component, $d=1$, i.e., just a linear combination of the predictors are sufficient for the regression of $Y$ on $X$. Following the  PLS methodology, 
\cite{cook2013envelopes} and  \cite{cook2021pls} give
 an estimator of that linear combination using $\widehat{\beta} X$ with $\widehat{\beta} $ the first Partial Least Square component.
Using linear regression after reduction, they showed that for this example (even with small $p$), the standard errors coefficient of the OLS fit are 
 about 6 and 60 times those of the PLS estimator.  
These types of differences are not unusual in the envelope (PLS) literature, as illustrated in examples from \cite{cook2018introduction}.
We added here the non parametric prediction estimator using all four variables and the non parametric prediction estimator 
after reduction to $\widehat{\beta} X$. 
From left to right, Figure \ref{fig:musselsfit} shows the prediction and confidence intervals using linear reduction, (as in  \cite{cook2018introduction}, \cite{cook2023libro}), using non parametric regression after reduction  $\widehat{\beta} X$,  and non parametric regression with all the predictors. 
We found again that even if the dimension of the predictors is small, doing 
reduction makes the confidence intervals much smaller and non-parametric and linear regression gives similar confidence intervals. Here it is a case where linear regression is the natural choice nevertheless the importance of this example is to show that the reduction steps makes gain a lot of efficiency in the prediction.

% {\color{red}{Conclusions??? creo que no. pero no se}}
 
\section{Appendix}

\subsection{Proof of Theorem \ref{asymptotic}}

Recall that  $\bW=\bbe_0\bX$,    $\bw_0=\bbe_0\bx_0$ and $\regd(\bw)=\esp(Y\mid \bW=\bw)$. 
In order to prove Theorem \ref{asymptotic}
we need two results. The first one, Theorem \ref{EL_teo}, establishes the  asymptotic result  for $\widehat\regd(\bw_0)$, defined in (\ref{est-lim1}), which  was proven originally in \cite{nadaraya1964estimating},
\cite{nadaraya1964estimating},  \cite{parzen1962estimation}. 
For the sake of completeness, we include  a proof of  its  consistency and of its asymptotic normality, following the ideas  presented in  \cite{ziegler2003asymptotic}  for the Nadaraya–Watson  estimator when $d=1$. 
The second result, Proposition \ref{ktaylor}, is needed to prove the asymptotic equivalence between $\widehat\regd(\bw_0)$ and $\widehat\eta(\xx_0)$:
\begin{equation}\label{recreo2}
	\sqrt{ nh^d_n} \{\widehat\regd(\bw_0)-\widehat \eta(\xx_0)\}=o_P(1)\;.
\end{equation}

\begin{theorem}
	\label{EL_teo}
	Under conditions K.1, A.1-A.3 and if  $h_n\to 0$,  $nh^d_n\to \infty$ as $n\to \infty$ 	
\begin{equation}
	\label{dos_consistencias}
	\frac{1}{nh^d_n}\displaystyle{\sum_{i=1}^n} Y_i^a K\left(\frac{\bw_0-\bW_i}{h_n}\right)\convprob \regd^a(\bw_0)f_\bW(\bw_0)\;,\quad \hbox{for $a=0,1$}\; .
\end{equation}

	Moreover,  if  A.4, K.2 and $nh_n^{d+2q}\to 0$ also hold, then
	\begin{equation}\label{recreo1}
		\sqrt{ nh^d_n} \{\widehat \regd(\bw_0)-\regd(\bw_0)\}\convdist \mathcal N\left(0, \frac{\sigma^2(\bw_0)  \int_{\mathbb R^d} K^2(\bu)\;d\bu}{f_\bW (\bw_0)}\right)\;,
	\end{equation}
	
\end{theorem}

\begin{proposition} Under the same conditions of Theorem \ref{asymptotic}, for $a=0,1$,
	\label{ktaylor}
	\begin{equation}
		\label{each}
		\sqrt{nh^d_n}\frac{1}{nh^ d}	\sum_{i=1}^n  Y^a_i \left\{ K\left(\frac{ \bbe_0\xx_0-\bbe_0\XX_i}{h_n}\right)-
		K\left(\frac{ \widehat \bbe_n\xx_0-\widehat\bbe_n\XX_i}{h_n}\right)\right\}=o_P(1).
	\end{equation}
\end{proposition}
\subsubsection{Proof of Theorem \ref{asymptotic}}

Note that 
\begin{eqnarray*}
	\sqrt{ nh^d_n} \{\widehat \eta(\xx_0)-\eta(\xx_0)\}&=&
	\sqrt{ nh^d_n} \{\widehat \eta(\xx_0)-\widehat\regd(\bw_0)\}+
	\sqrt{ nh^d_n} \{\widehat\regd(\bw_0)-\regd(\bw_0)\}.
\end{eqnarray*}

Then,  \eqref{dist_asint} follows from Theorem \ref{EL_teo} as far as we show that the second term in the previous display  converges to zero. 
To prove so, write

\begin{align*}	
& \sqrt{ nh^d_n} \{\widehat\regd(\bw_0)-\widehat \eta(\xx_0)\}=	\sqrt{nh_n^d}\left\{\frac{\widehat A_n^{ (1)}}{\widehat A_n^{ (0)}}-\frac{ A_n^{ (1)}}{ A_n^{ (0)}}\right\}\\
	 &=\frac{1}{\widehat A_n^{ (0)} A_n^{ (0)}}\left\{\widehat A_n^{ (1)} \;\sqrt{nh_n^d}(\widehat A_n^{ (0)}- A_n^{ (0)})-\;\sqrt{nh_n^d}(\widehat A_n^{ (1)}-A_n^{ (1)}) \widehat A_n^{ (0)}\right\},
\end{align*}
where for $a=0,1$,
$$\widehat A_n^{(a)}= \frac{1}{nh^d_n}\sum_{i=1}^ n Y^a_i  K\left( \frac{{\widehat\bbe_n}  (\xx_0 - \XX_i)}{h_n}\right)\;,\;\;\ A_n^{(a)}= \frac{1}{nh^d_n}\sum_{i=1}^ n Y^a_i  K\left( \frac{\bbe_0  (\xx_0 - \XX_i)}{h_n}\right).$$
Combining (\ref{each}) with  (\ref{dos_consistencias}), we get that 
\begin{equation}
	\label{A_n_sombrero}
	\widehat A^{(a)}_n\convprob \regd^a(\bw_0)f_\bW(\bw_0)\;, \;\;  A^{(a)}_n\convprob \regd^a(\bw_0)f_\bW(\bw_0)\;,
	\end{equation}
and, as a consequence, since $f_\bW(\bw_0)>0$, $\sqrt{ nh^d_n} \{\widehat\regd(\bw_0)-\widehat \eta(\xx_0)\}=o_P(1).$  
The consistency stated in \eqref{consistencia} is a direct consequence of \eqref{dist_asint}.

\subsubsection{Proof of Theorem \ref{EL_teo}:}

The following result, which is a restatement of 
Parzen's (1962) proposition \cite{parzen1962estimation},
will be invoked many times during the proof of  Theorem \ref{EL_teo}.

\begin{proposition}
	\label{parzen}
	Let $L: \mathbb R^ d\to \mathbb R$ be  bounded integrable  with  $\displaystyle{\lim_{\Vert \bz\Vert \to \infty} \Vert\bz L(\bz)\Vert =0}$. Let $\bW$ be a random vector with density function $f_\bW$. Then, if  $\esp\{\psi(\bW)\}<\infty$
	\begin{equation}
		\frac{1}{h^d_n}\esp\left\{\psi(\bW)L\left(\frac{\bw-\bW}{h_n}\right)\right\}\to \psi(\bw)f_W(\bw)\;\int_{\mathbb R^ d }L(\bu)d\bu\;,\quad \hbox{when $h\to 0$}, 
	\end{equation}
	for every continuity point $\bw$ of $\Psi$ and $f_\bW$. 
	
\end{proposition}

%\textcolor{red} {Este es un resultado clave para las cuentas que estamos haciendo. En mi epoca, se estudiaba en Real. Seria bueno ver en que parte de la carrera esta incluido. Podria darse, incluso, en avanzado. Pero tiene un millon de aplicaciones en todo lo que tenga que ver con nucleo. }
%\section*{Parzen's (1962)lemma}
%\begin{proposition} \textbf{Parzen's (1962)lemma}
%	
%	
%	Let $L:\mathbb R^d\to \mathbb R$ be bounded  integrable with ${\lim_{\Vert \bz\Vert \to \infty }\vert \bz L(\bz)\vert=0}$ and let $\phi: \mathbb R^ d\to \mathbb R$ be intergrable. Then 
%	\begin{equation}
%		\lim_{h\to 0}\frac{1}{h^d_n}\int _{\mathbb R^ d} L\left(\frac{\xx-\bz}{h_n}\right) \phi(\bz)  d\bz =\phi(\xx)\int_{\mathbb R^ d}   L(z) dz.
%	\end{equation}
%	for every continuity point $x$ of $\phi$.
%	
%\end{proposition}
%

Now, to prove each of the convergences stated in \ref{dos_consistencias}, we will show that the estimators  are asymptotically unbiased and that their variances converge to zero.

\noindent {\bf  Asymptotically unbiased: }
The continuity of $\regd^a$ and $f_\bW$ at $\bw_0$ assumed in A.3 and A.1, respectively,  combined with condition K.1, allow to invoke   Proposition \ref{parzen} with $L=K$ and $\psi=\regd^ a$ since, by A.2 $\esp(\vert Y\vert)<\infty$, to conclude that since $h_n\to 0$
\begin{eqnarray*}	
	\esp \left\{\frac{1}{nh^d_n}\sum_{i=1}^n Y_i^a K\left(\frac{\bw_0-\bW_i}{h_n}\right)\right\}
	&\to &
	\regd^ a(\bw_0)f_\bW(\bw_0).
\end{eqnarray*}

\noindent {\bf Variance converges to 0: }
%Consider
%Using  that $K$ is bounded, by {\it E.1}, we get that
%\begin{eqnarray*}
%	\var \left\{\frac{1}{nh^d_n}\sum_{i=1}^n Y_i^a K\left(\frac{\bw_0-\bW_i}{h_n}\right)\right\}&\leq&
%	\frac{1}{nh^{2d}}
%	\esp \left\{ Y^{2a} K^2\left(\frac{\bw_0-\bW}{h_n}\right)\right\}  \label{recreo4} \\ &= &
%	\frac{1}{nh^{2d}}\esp \left\{ \regd_2^a(\bW)    K^2\left(\frac{\bw_0-\bW}{h_n}\right)\right\} \nonumber  \;.
%\end{eqnarray*}
\begin{align*}
	\var \left\{\frac{1}{nh^d_n}\sum_{i=1}^n Y_i^a K\left(\frac{\bw_0-\bW_i}{h_n}\right)\right\}&\leq
	\frac{1}{nh_n^{2d}}
	\esp \left\{ Y^{2a} K^2\left(\frac{\bw_0-\bW}{h_n}\right)\right\}  \label{recreo4} \\
	&=
	\frac{1}{nh_n^{d}}\frac{1}{h^d_n}\esp \left\{ \regd_2^a(\bW)    K^2\left(\frac{\bw_0-\bW}{h_n}\right)\right\} \nonumber  \;.
\end{align*}
Using A.1-A.3 and K.1, we can invoke 
Proposition \ref{parzen} with  $L=K^2$, $\psi=\regd^a$  to show that
the second factor converges when $h_n\to 0$ and therefore, the variance considered converges to 0 since  $nh_n^ d$ goes to infinity.

\noindent{\bf Proof of  the asymptotic distribution \eqref{recreo1}. }

To show the asymptotic normality presented in display (\ref{recreo1}), note that
\begin{eqnarray}
	\widehat\regd (\bw_0)-\regd(\bw_0)&=&
		\frac{1}{\widehat f_\bW(\bw_0)}	\frac{1}{nh^d_n}\;\sum_{i=1}^n \left\{Y_i -\regd(\bw_0)\right\}K\left(\frac{\bw_0-\bW_i}{h_n}\right)\;,
\end{eqnarray}
where 
$$
\widehat f_\bW(\bw_0)=\frac{1}{nh^d_n}\sum_{i=1}^n K\left(\frac{\bw_0-\bW_i}{h_n}\right)\convprob f_\bW(\bw_0)\;, $$
as we established in (\ref{dos_consistencias}).
Therefore, by Slutsky,  we need to  prove that 
\begin{equation}
	\frac{\sqrt{nh^d_n}}{nh^d_n}\sum_{i=1}^n \!  \left\{Y_i -\regd(\bw_0)\right\}K \! \! \left(\frac{\bw_0-\bW_i}{h_n}\right)\! \!  \convdist \!  \mathcal N\left(0, \sigma^2(\bw_0) f_\bW (\bw_0) \! \int_{\mathbb R^d} \! \! \!  K^2(\bu) d\bu\right) \label{recreo10}.
\end{equation}
Since $nh_n^{d+2q}\to 0$, by   K.2,  this result is a consequence of decomposing 
\begin{equation}
	\label{lostres}
	\frac{\sqrt{nh^d_n}}{nh^d_n}\sum_{i=1}^n \left\{Y_i -\regd(\bw_0)\right\}K\left(\frac{\bw_0-\bW_i}{h_n}\right)=A_n(\bw_0)+\sqrt{nh_n^{d+2q}}b_n(\bw_0)+o_P(1)\;,
\end{equation}
where 
\begin{eqnarray}
	\label{An}
	A_n(\bw_0)&=&\frac{\sqrt{nh^d_n}}{nh^d_n}\sum_{i=1}^n \left\{Y_i -\regd(\bW_i)\right\}K\left(\frac{\bw_0-\bW_i}{h_n}\right) \nonumber\\
	&&\convdist  \mathcal N\left(0, \sigma^2(\bw_0) f_\bW (\bw_0) \int_{\mathbb R^d} K^2(\bu)\;d\bu\right)\;,
\end{eqnarray}
$$b_n(\bw_0)\to \sum_{\vert{\balfa}\vert =q}
\frac{(-1)^{q}}{q!} \frac{\partial^{\balfa} \regden(\bw_0)}{\partial \bu^{\balfa}}\int_{\mathcal R^d}\bv^{\balfa} K(\bv) \;d\bv.$$

To show (\ref{lostres}), note that (adding and substracting $\regd(\bW_i)$) and calling $D_n(\bw_0)=B_n(\bw_0)+R_n(\bw_0)$,
%$$
%\frac{\sqrt{nh^d_n}}{nh^d_n}\sum_{i=1}^n \left\{Y_i -\regd(\bw_0)\right\}K\left(\frac{\bw_0-\bW_i}{h_n}\right)=A_n(\bw_0)+B_n(\bw_0)+R_n(\bw_0)\;,$$
\begin{eqnarray*}
\frac{\sqrt{nh^d_n}}{nh^d_n}\sum_{i=1}^n \left\{Y_i -\regd(\bw_0)\right\}K\left(\frac{\bw_0-\bW_i}{h_n}\right)&=&A_n(\bw_0)+B_n(\bw_0)+R_n(\bw_0),\end{eqnarray*}
where $
D_n(\bw_0)=\frac{\sqrt{nh^d_n}}{nh^d_n}\sum_{i=1}^n \left\{\regd(\bW_i) -\regd(\bw_0)\right\}K\left(\frac{\bw_0-\bW_i}{h_n}\right),$ and $B_n(\bw_0)=\esp\{D_n(\bw_0)\}$,
stands for the asymptotic bias  while $R_n(\bw_0)$ captures the remaining variability, which we will show  converges to zero. Namely, 
\begin{equation} 	R_n(\bw_0) =D_n(\bw_0
)-B_n(\bw_0)\convprob 0\;.
\end{equation}

Let start with the bias term $B_n(\bw_0)=\esp\{D_n(\bw_0)\}$, calling $\regden(\bu)=\{\regd(\bu)-\regd(\bw_0)\} f_\bW(\bu)$.
\begin{eqnarray*}
\esp\{D_n(\bw_0)\}
%=
%\frac{\sqrt{nh^d_n}}{h^d_n}\esp\left\{\left\{\regd(\bW) -\regd(\bw_0)\right\}K\left(\frac{\bw_0-\bW}{h_n}\right)\right\}
=\sqrt{nh^d_n}\int_{\mathcal R^d}\regden (\bw_0-h\bv) K\left(\bv\right)\;d\bv,
\end{eqnarray*}
%	\begin{eqnarray}
%		\esp(B_n(\bw))&=&\frac{\sqrt{nh^d_n}}{h^d_n}\esp\left\{\left\{\regd(\bW) -\regd(\bw)\right\}K\left(\frac{\bw_0-\bW}{h_n}\right)\right\}\nonumber \\ &=&
%		\frac{\sqrt{nh^d_n}}{h^d_n}\int_{\mathcal R^d}\left\{\regd(\bu) -\regd(\bw)\right\}K\left(\frac{\bw_0-\bu}{h_n}\right)f_\bW(u)\;du
%		\nonumber \\&=&\sqrt{nh^d_n}\left\{\frac{1}{h^d_n}\int_{\mathcal R^d}\regden (\bu) K\left(\frac{\bw_0-\bu}{h_n}\right)\;d\bu\right\}\label{recreo20}
%	\end{eqnarray}
%
%Parzen's (1962) lemma states that 
%\begin{equation}
%	\label{indet}
%	\frac{1}{h^d_n}\int_{\mathcal R^d}\regden (\bu) K\left(\frac{\bw_0-\bu}{h_n}\right)\;d\bu \to \regden(\bw_0)=0\;.
%\end{equation}
%On the other hand, $\sqrt{nh^d_n}\to \infty$. Thus, at this point, we need to solve an indetermination.
%\textcolor{red}{Aca viene Taylor. grado que estamos eligiendo es apenas para que sea compatible con las hipotesis que pedimos en la convergencia uniforme. En un principio, podemos aca dejar hecho un desarrollo  ad hoc, que dependa de la regularidad de la funcion que estamos expandiendo. La idea es que se entienda que es lo minimo que se pide para la distribucion asintotica del estimador no parametrico. Despues aparecen condiciones adicionales  para tener compatibilidad con la convergencia uniforme y los procesos empiricos. }
%\textcolor{blue}{Dime que resultado quieres y te dire que tasa necesitas. }
%Changing variables on the integral presented in display (\ref{indet}), 
Using the fact that $\regden (\bw_0) =0$ 
and a Taylor expansion until order $q$  for $\regden$ around $\bw_0$, allowed by assumption A.4, we have that, for $c_\bv\in (0,1)$,
\begin{eqnarray}
\nonumber\esp\{D_n(\bw_0)\}&=&\sqrt{nh^d_n} \int_{\mathcal R^d}\regden (\bw_0 - h\bv) K(\bv) \;d\bv\\
&=& \label{matosesgo}\sqrt{nh^d_n} \sum_{\vert\balfa\vert=1}^{q-1} (-h)^{\vert\balfa\vert } \frac{1}{\balfa!} \frac{\partial^{\balfa} \regden(\bw_0)}{\partial \bu^{\balfa}}\int_{\mathcal R^d}\bv^{\balfa} K(\bv) \;d\bv\\
& +&   \nonumber\sqrt{nh^d_n} \sum_{\vert{\balfa}\vert =q}
\frac{(-h)^{q}}{q!} \int_{\mathcal R^d}\frac{\partial^{\balfa} \regden(\bw_0+c_\bv h \bv)}{\partial \bu^{\balfa}}\bv^{\balfa} K(\bv) \;d\bv ,	
\end{eqnarray} 
where  $\bv^{\balfa}=(v_1^{\alpha_1}, \dots, v_r^{\alpha_r})$, with $\balfa=(\alpha_1,\dots, \alpha_r) $ and $\bv = (v_1,\dots, v_r)$. By assumption K.2, the term presented in  (\ref{matosesgo}) is equal to zero and, therefore, we conclude that $B_n(\bw_0)=\sqrt{nh^{d+2q}}b_n(\bw_0)$,
where, by A.4 and K.3, we get that  
$$b_n(\bw_0)\to \sum_{\vert{\balfa}\vert =q}
\frac{(-1)^{q}}{q!} \frac{\partial^{\balfa} \regden(\bw_0)}{\partial \bu^{\balfa}}\int_{\mathcal R^d}\bv^{\balfa} K(\bv) \;d\bv \;.$$
%\textcolor{red}{ver que queda aca y debatir por que es que en este trabajo nos importa quedarnos con este orden. yo aun no lo entendi. veo que es por algo de la convergencia uniforme. pero aun  no cacho bien como juntar todo.}
Recall now that $R_n(\bw_0)=D_n(\bw_0)-\esp\{D_n(\bw_0)\}$ are centered random variables with 
\begin{equation}
\var(R_n(\bw_0))=	\var(D_n(\bw_0))\leq 	\frac{1}{h^d_n}\esp\left(\left\{\regd(\bW) -\regd(\bw_0)\right\}^ 2 K^ 2\left(\frac{\bw_0-\bW}{h_n}\right)\right)\;,
\end{equation}
that goes to 0, invoking Proposition \ref{parzen} with $L=K^2$ and $\psi=\{\regd-\regd(\bw_0)\}^2$. 
% the left hand side converges, up to a constant factor that depends on $\bw_0$,  to $\regden(\bw_0)=0$.
%\textcolor{red}{necesitamos que $ \left\{\regd(\cdot) -\regd(\bw_0)\right\}^ 2f_\bW(\cdot)$ sea integrable y continua en $\bw_0$. }

\noindent {\bf Proof of (\ref{An})}
The proof of  this asymptotic behaviour, is a consequence of the Lindeberg Theorem for triangular arrays  applied to 
\begin{equation}
	T_{n,i}= \frac{1}{\sqrt{n s_n^ 2}}\left\{Y_i -\regd(\bW_i)\right\}K\left(\frac{\bW_i-\bw}{h_n}\right)\;
\end{equation}
with $s_n^ 2=\var\left(\left\{Y_i -\regd(\bW_i)\right\}K\left(\frac{\bw_0-\bW_i}{h_n}\right) \right)$.
 To fit the Lindeberg  framework, note that
$$\esp\left(\left\{Y_i -\regd(\bW_i)\right\}K\left(\frac{\bw_0-\bW_i}{h_n}\right) \right)=0, $$
and that the Lyapounov condition, which actually implies Lindeberg's, 
\begin{equation}
	\label{lili_te_quiero}\lim_{n\to \infty}\sum_{i=1}^ n \esp(\vert T_{n,i}\vert ^ {2+\delta})=0.
	\end{equation}
To prove \eqref{lili_te_quiero}
we will show that, under condition A.2
$$\lim_{n\to \infty}\sum_{i=1}^ n \esp(\vert T_{n,i}\vert ^ {2+\delta})=0$$
To do so, note that
\begin{eqnarray*}
\sum_{i=1}^ n \esp(\vert T_{n,i}\vert ^ {2+\delta})=
\frac{ (nh^ d)^{-{\delta/2}}}{(s_n^2/h^d_n)^{1+\delta/2}} \; \frac{1}{h^d_n}\esp\left(\vert Y -\regd(\bW)\vert^ {2+\delta} \left\vert K \left(\frac{\bw_0-\bW}{h_n}\right)\right\vert ^ {2+\delta}\right).
\end{eqnarray*}
A.2, A.3 and K.1 allow us to invoke Proposition \ref{parzen} again  with $\psi=\sigma^{2+\delta}$ defined in \eqref{regd} and $L=K^{2+\delta}$, to conclude that the second factor is convergent. The first factor goes to zero, since $nh_n^ d\to \infty$ and
\begin{eqnarray}
	\label{var_asin}
\nonumber	s_n^ 2&=&	\esp\left(\left\{Y_i -\regd(\bW_i)\right\}^ 2K^2\left(\frac{\bW_i-\bw}{h_n}\right) \right)\\
	&=&\esp\left(\sigma^2(\bW_i)K^2\left(\frac{\bW_i-\bw}{h_n}\right) \right)=
	h_n^ d \sigma^2(\bw_0)f_\bW(\bw_0)\int K^ 2(\bu)d\bu +o(h^d_n),
\end{eqnarray}
by Proposition \ref{parzen} with $\psi=\sigma^ 2$ and $L=K^ 2$. 
Therefore, the Lindeberg's theorem guarantees that $\sum_{i=1} ^ n T_{n,i}$ converges in distribution to a standard Gaussian law. 
Finally,  from  (\ref{var_asin}) and Slutzky, we conclude that 
$$A_n(\bw_0)=\frac{1}{\sqrt{nh^d_n}}\sum_{i=1}^ n 	\left\{Y_i -\regd(\bW_i)\right\}K\left(\frac{\bW_i-\bw_0}{h_n}\right) \to
\mathcal N\left(0,\sigma^2(\bw_0)f_\bW(\bw_0)\int K^2(\bu)d\bu\right)\;. $$
This concludes the proof of the asymptotic normality stated in (\ref{recreo10}).
%
%\begin{eqnarray}
%	\label{var_asin}
%	s_n^ 2&=&\var\left(\left\{Y_i -\regd(\bW_i)\right\}K\left(\frac{\bw_0-\bW_i}{h_n}\right) \right)=	\esp\left(\left\{Y_i -\regd(\bW_i)\right\}^ 2K^2\left(\frac{\bW_i-\bw}{h_n}\right) \right)\\
%	&=&
%	h_n^ d \sigma^2(\bw_0)f_\bW(\bw_0)\int K^ 2(\bu)d\bu +o(h^d_n),
%\end{eqnarray}
%
%
%
%\begin{equation}
%	\label{var_asin}
%	s_n^ 2=\var\left(\left\{Y_i -\regd(\bW_i)\right\}K\left(\frac{\bw_0-\bW_i}{h_n}\right) \right)=	%\esp\left(\left\{Y_i -\regd(W_i)\right\}^ 2K^2\left(\frac{W_i-w}{h_n}\right) \right)=
%	h_n^ d \sigma^2(\bw_0)f_\bW(\bw_0)\int K^ 2(\bu)d\bu +o(h^d_n),
%\end{equation}
%as a consequence of Proposition \ref{parzen}. 

\subsubsection{Proof of Proposition \ref{ktaylor}}
By condition K.3, we can apply a Taylor expansion to $K(\bu)=k(\Vert\bu\Vert)$ to get
\begin{eqnarray}
&&				\sqrt{nh^d_n}\frac{1}{nh^ d} \sum_{i=1}^n Y_i^a \left\{  K\left( \frac{\widehat{\bbe}_n (\xx_0 - \XX_i)}{h_n}\right) -    K\left( \frac{{\bbe_0}  (\xx_0 - \XX_i)}{h_n}\right) \right\}\\
&=& 
\label{bueno}\sqrt{nh^d_n}\frac{1}{nh^ {d+1}} \sum_{i=1}^n Y_i^a g\left(\frac{{\bbe}_0  (\xx_0 - \XX_i)}{h_n}\right)
M ( \xx_0-\XX_i )\hbox{vec} (\widehat{\bbe}_n - \bbe_0)  \\
&+& 
\label{resto}  \frac{\sqrt{nh^d_n}}{nh^ {d+2}}\hbox{vec}^T\! (\widehat{\bbe}_n\! - \! \bbe_0) \!\!\sum_{i=1}^n \!Y_i^a \! M(\xx_0\! -\! \XX_i)^ TR_i M(\xx_0-\XX_i) \hbox{vec} (\widehat{\bbe}_n \!  - \!  \bbe_0),
\end{eqnarray}
for $g:\mathbb R^ d\to \mathbb R^d$ and $M: \mathbb R ^{p}\to \mathbb R^ {d\times pd }$ defined as
\begin{equation}
\label{gM}
g(\bw) =k' \left(\left\Vert \bw \right\Vert\right) \frac{\bw^ T}
{\left\Vert \bw\right\Vert}\;,\quad M(\bx) := \bx^T \otimes I_d ,
\end{equation}
and  remainder  terms  $R_i$ given by
\begin{eqnarray*}
R_i&=&\frac{1}{2}k'\left(\left\Vert\frac{\widetilde{\bbe}  (\xx_0 - \XX_i)}{h_n}\right\Vert\right)\left\{\frac{h_n I_d}{\Vert
\widetilde{\bbe}  (\xx_0 - \XX_i)\Vert}-\frac{h_n \widetilde{\bbe}(\xx_0 - \XX_i)\left\{\widetilde{\bbe} (\xx_0 - \XX_i)\right\}^T}{\Vert\widetilde{\bbe} (\xx_0 - \XX_i)\Vert^3} \right\}\\
&+& \frac{1}{2}k''\left(\left\Vert\frac{\widetilde{\bbe} (\xx_0 - \XX_i)}{h_n}\right\Vert\right)
\frac{\widetilde{\bbe}  (\xx_0 - \XX_i)\left\{\widetilde{\bbe} (\xx_0 - \XX_i)\right\}^T }{\Vert \widetilde{\bbe}  (\xx_0 - \XX_i)\Vert ^2},
\end{eqnarray*}
where $\widetilde \bbe$ lies between $\bbe_0$ and $\widehat \bbe_n$. 
With these definitions, since  $\sqrt{n}(\widehat{\bbe}_n - \bbe_0) =O_p(1)$, (\ref{each}) will be a  consequence of the following convergences, which will be proven in what follows. 
\begin{equation}
	\label{primera} \frac{\sqrt{h^d_n}}{h^d_n} \esp \left( \frac{1}{h_n} Y^a g \left(\frac{{\bbe}_0  (\xx_0 - \XX)}{h_n} \right)
M  (\xx_0-\XX)
\right)  \rightarrow   0 
\end{equation}
\begin{equation}
\frac{1}{n h^d_n} \esp\left( \frac{1}{h^2_n} Y^{2a}  \! \left( \!
g \left(\frac{{\bbe}_0  (\xx_0 \!-  \!\XX)}{h_n}  \!\right)
M  (\xx_0 \!- \!\XX) \!
\right)^T \! \!
g  \!\left( \!\frac{{\bbe}_0  (\xx_0 \! -  \!\XX)}{h_n}  \!\right) \!
M  \! (\xx_0 \!- \!\XX) \!\right) \!
 \rightarrow   \! 0 \label{varianza1}
 \end{equation}
 \begin{equation}
\frac{1}{\sqrt{n h^{d+4}}}  \frac{1}{n}   \sum_{i=1}^n Y_i^a M(\xx_0-\XX_i)^ TR_i M(\xx_0-\XX_i)      \rightarrow    0.  \label{resto}
\end{equation}
%where $\bar{R} $ indicate the mean of $R$.

\noindent {\bf Proof of  (\ref{primera})}
Recall that $\bW=\bbe_0\XX$. To simplify the notation, both for $a=0,1$, we use $L(\bw)$ to denote the function defined by 
\begin{eqnarray*}
\tilde M(\bw)&=&   \esp \left(Y^a  
M  (\xx_0-\XX)\mid {\mathbf W} = \bw\right) \in {\mathbb R^{1\times pd}}, \hbox{ with } \bw \in {\mathbb R}^d.
\end{eqnarray*}
Conditioning on $\bW$,
%and recalling that 
%$f_{\bbe_0\XX}$ stands for the density of $\bbe_0\XX$
we get that 
\begin{eqnarray*}
\frac{1}{h^d_n}\esp\left( Y^a g\left(\frac{{\bbe}_0  (\xx_0 - \XX)}{h_n}\right)
M ( \xx_0-\XX)\right)&=&\frac{1}{h^d_n}\esp\left(  g\left(\frac{{\bbe}_0\xx_0 -\bbe_0 \XX}{h_n}\right)
\tilde M(\bbe_0\XX)\right).
%\\
%&=&\frac{1}{h_n}\int_{\mathbb R^ d}  g\left(\frac{{\bbe}_0 \xx_0 - \bz)}{h_n}\right)L(\bz)
%f_{\bbe_0\XX}(\bz) \; d\bz.
\end{eqnarray*}

%
%Invoking Proposition \ref{parzen}  to each coordinate of $g$ in  lieu of $L$, taking  $\psi=\tilde M$ and recalling that A.5 and  K.3 hold, we conclude that  
%$$\frac{1}{h^d_n}\esp\left(  g\left(\frac{{\bbe}_0\xx_0 -\bbe_0 \XX}{h_n}\right)
%\widetilde M (\bbe_0\XX)\right)\to \left\{\int_{\mathbb R^d}   g\left( 
%\bu  \right) d\bu\right\}\; \tilde M(\bbe_0\xx_0)f_{\bbe_0\XX}(\bbe_0\xx_0) =0\;,$$
%
%$$\frac{1}{h^ d_n} \int_{\mathbb R^d}   g\left( 
% \frac{\bbe_0 \xx_0-\bz}{h_n} \right) r(\bz)  \; d\bz\to r(\bbe_0\xx_0) \int_{\mathbb R^d}   g\left(
% \bu  \right) d\bu=0\;,$$
%since, by condition  K.3 , the integral of each coordinate of $g$ over $\mathbb R^ d$  vanishes. Moreover, we will  prove that the left hand side of the previous display is of order $O(h)$. In fact,   
Defining $r : {\mathbb R}^d \rightarrow {\mathbb R}^{pd}$ as $r(\bu) = \widetilde M (\bu) f_{\bW} (\bu) $,  note that, using K.3 and A.5,
\begin{align*}
& \frac{1}{h^{d+1}_n}\esp\left(  g\left(\frac{{\bbe}_0\xx_0 -\bbe_0 \XX}{h_n}\right)
\widetilde M (\bbe_0\XX)\right)=
\frac{1}{h^ {d+1}_n} \int_{\mathbb R^d}   g\left( 
\frac{\bw_0-\bz}{h_n} \right) r(\bz)  \; d\bz\\
&= \int_{\hbox{sup}(K)}   g\left( 
\bu \right) \frac{1}{h_n}\left\{r(\bw_0+h\bu) -r(\bw_0)\right\} \; d\bu \to \nabla r(\bw_0)\int _{\hbox{sup}(K)} g(\bu)\bu  \;d\bu\;,
\end{align*}
where $\hbox{sup}(K)$ stands for the support of $K$, concluding \eqref{primera} holds. 

\noindent {\bf Proof of (\ref{varianza1})}
Recalling  the definition of $g$ and $M$ presented in (\ref{gM}), the left  hand side of display (\ref{varianza1}) can be written as
\begin{eqnarray*}
\esp\left(  Y^{2a}
\left\{k' \left(\frac{\Vert{\bbe}_0  (\xx_0 - \XX)\Vert}{h_n} \right)\right\}^ 2
\Vert \xx_0-\XX\Vert ^2\right)=  \esp\left(
\left\{k' \left(\frac{\bw_0 - \bW}{h_n} \right)\right\}^ 2L_2(\bW) \right),
\end{eqnarray*}
where, for $ \bw \in {\mathbb R}^d$, $L_2(\bw)=   \esp \left(Y^{2a}  
\Vert \xx_0-\XX\Vert ^ 2\mid {\mathbf W}= \bw\right)$. 
%\begin{eqnarray*}
%L_2(\bw)&=&   \esp \left(Y^{2a}  
%\Vert \xx_0-\XX\Vert ^ 2\mid {\mathbf W}= \bw\right),  \hbox{ for } \bw \in {\mathbb R}^d. 
%\end{eqnarray*}
K.3 and the assumed regularity established in A.5, allows us to use Proposition \ref{parzen} with $\psi=L_2 $ and $L=(k')^2$ to conclude that 
%As in the proof of (\ref{esperanza1}), defining now  $\tilde g(\bw)=\{k^\prime(\bw)\}^2$ and $\tilde r(\bw)=L_2(\bw)f_\bW(\bw)$, we get that 
\begin{eqnarray*}
\frac{1}{ h_n^{d}} \esp\left(
	\left\{k' \left(\frac{\bw_0 - \bW}{h_n} \right)\right\}^ 2L_2(\bW) \right) 
%	\frac{1}{h^ {d+1}_n} \int_{\mathbb R^d}   \tilde g\left( 
%	\frac{\bw_0-\bz}{h_n} \right) \tilde r(\bz)  \; d\bz\\
%	&=& \int_{\hbox{sup}(K)}  \tilde  g\left( 
%	\bu \right) \frac{1}{h_n}\left\{\tilde r(\bw_0+h\bu) -\tilde r(\bw_0)\right\} \; d\bu\\
%	&\to& \nabla \tilde r(\bw_0)\int _{\hbox{sup}(K)} \tilde g(\bu)\bu  \;d\bu\;,
		&\to& \left\{ \int_{\mathbb R^d} \{k' (\bu ) \}^2 \right\} L_2(\bw_0) f_{\bw}(\bw_0).
\end{eqnarray*}
%\textcolor{red}{ver que hay que pedirle a $r$}
%under K.3 and  A.5. 
Therefore, (\ref{varianza1}) holds since 
$nh_n^2\to \infty$.

\noindent {\bf Proof of (\ref{resto})}
By K.3 and K.4, $\Vert R_i\Vert \leq C$ and therefore
$$  \left\Vert \frac{1}{n}   \sum_{i=1}^n Y_i^a M(\xx_0-\XX_i)^ TR_i M(\xx_0-\XX_i)\right\Vert \le C  \frac{1}{n}   \sum_{i=1}^n \vert Y_i^a\vert  \left\Vert \xx_0-\XX_i\right\Vert^ 2=O_P(1)$$
by A.5 and the Law of Large Numbers. Finally, since  $n h_n^{d+4} \rightarrow 0$,  (\ref{resto}) holds.

\subsection{Proof of  Theorem \ref{consistency}}

%As in \eqref{A_n_sombrero}, from Theorem \ref{EL_teo} and Lema \ref{ktaylor}, we get that 
%$$
%\left\vert \frac{1}{nh_n^d}\sum_{i=1}^ nY_i^a K\left(\frac{\widehat \bbe_n (\xx_0-\XX)}{h_n}\right)-\regd^a(\bbe_0\xx_0)f_\bW(\bbe_0\xx_0)\right\vert\convprob   0.
%$$
%In this section we will show that this convergence is preserved when taking the supreme with $\xx_0$ varying in a compact set $\mathcal K$. To do so, 
Following the empirical process literature, we use the abbreviation $Qf=\int f dQ$ and   $\PP_n$  for the empirical measure, being the linear combination $\PP_n=n^{-1}\sum_{i=1}^n\delta_{(\XX_i,Y_i)}$ of the Dirac measures at the observarions $(\XX_i,Y_i)$.  Thus, $\PP f$ stands for $\esp f(\XX,Y)$ when  $(\XX, Y)\sim \PP$ while $\PP_nf=n^{-1}
\sum_{i=1}^n f(\XX_i, Y_i)$.

For $a\in \{0,1\}$, $\xx_0 \in \mathbb{R}^{p}$ and $\bbe \in \mathbb{R}^{p\times d}$, define  
\begin{eqnarray}
	g^{\tiny (a)}_{ \tiny{{{\beta}}},\xx_0,h}(\xx,y)  &=& Y^{\tiny{a}} K\left( \frac{{\bbe} (\xx_0 - \xx)}{h_n}\right)\\
		 S_{a}(\bbe,\xx_0,h)&=&\esp\left(	g^{\tiny (a)}_{\tiny{\bbe},\xx_0,h}(\XX,Y) \right)=\frac{1}{h^d_n} \EE\left[Y^aK\left(\frac{\bbe (\xx_0-\XX)}{h_n}\right)\right]\\
		 S_{a}(\bbe,\xx_0)&=&\EE\left(Y^a\vert \bbe \XX=\bbe\xx_0\right)f(\bbe\xx_0, \bbe),
\end{eqnarray}
recalling that  $f(u,\bbe)$ denotes  the density function of  $\bbe\XX$ evaluated at $u$. 
According to this notation
\begin{equation*}	\frac{1}{h_n^d}\PP_n \gnt^{(\tiny a)}=\frac{1}{nh_n^d}\!\sum_{i=1}^ n \! Y_i^a K\left(\!\frac{\widehat \bbe_n (\xx_0\! -\!\XX_i)}{h_n}\!\right), \;
S_a(\bbe_0, \xx_0)=	\regd^a(\bbe_0\xx_0)f_\bW(\bbe_0\xx_0).
	\end{equation*}
We will be able to 
deduce the uniform  consistency of  $\widehat\eta(\xx_0)$, once we show that 
\begin{equation}
	\label{cada_uno}
\sup_{\xx_0\in \mathcal K}	\left\vert \frac{1}{h_n^d}\PP_n \gnt^{(\tiny a)}- S_a (\bbe_0,\xx_0)\right\vert\convprob 0\;,\quad a=0,1,
\end{equation}
%\begin{equation}
% \label{lapapa}
%\frac{1}{h_n^d}\PP_n \gnt^{(\tiny 0)}\cp f(\bbe_0^\trans\xx_0,\bbe_0)\quad \hbox{and}
%\quad \frac{1}{h_n^d}\PP_n \gnt^{(\tiny 1)} \cp \int y f(\bbe_0^\trans\xx_0,y,\bbe_0) dy\;,
%\end{equation}
uniformly in $\xx_0$ over compact sets.

The following results will be invoked in the proof of Theorem \ref{consistency}.

\begin{proposition}
	\label{bias} Consider a kernel $K$ that  satisfies S.1. Assume that $\widehat\bbe_n\convprob\bbe_0$ and  that 
	$h_n\to 0$. Then, for
 any compact set $\mathcal{K}\in \mathbb{R}^{p }$ we get that
	\begin{equation}
		\label{eqbias}
		\sup_{\xx_0 \in \mathcal{K}} \vert S_{a}(\widehat\bbe_n,\xx_0,h_n)- S_a(\bbe_0, \xx_0) \vert \convprob 0, 
	\end{equation}
provided that  for $a=0$ S.2 holds and for $a=1$ S.3 holds.
\end{proposition}

\textbf{Proof}
Since $K$ is bounded with compact support $\hbox{sup}(K)$  and  its integral equals 1,  we obtain that 
\begin{align*}
	&\sup_{\xx_0 \in \mathcal{K}} \vert  S_{0}(\widehat\bbe_n,\xx_0,h_n)- S_0(\bbe_0,\xx_0) \vert 
	\\
	&\leq \Vert K\Vert_\infty \int_{\hbox{sup}(K)}    \sup_{\xx_0 \in \mathcal{K}} \vert  
	f(\widehat\bbe_n\xx_0+h_n\bw,\widehat\bbe_n) -f(\bbe_0\xx_0, \bbe_0)\vert d\bw\;.
\end{align*}
Note that by S.2, $f(t,\bbe)$ is uniformly continuous over compact sets. Since $h_n \to 0$ and $K$ is bounded with compact support the integrand converges point wise to zero and the result follows from the Dominated Convergence Theorem. 
The same argument can be applied for $a=1$.   

\bigskip

\begin{proposition}
	\label{desig-max}
	Consider a kernel $K$ that  satisfies S.1. Assume that  $\esp(Y^2) <\infty$.  Then, for  some constant $M_a>0$,
	\begin{equation}
		\label{random}
		\EE \left[ \sup_{f\in \mathcal F_{n,a} } \vert \PP_n f - 
		\PP f\vert \right] \leq M_a \frac{1}{\sqrt n}\;,\quad \hbox{for $a=0,1$,}
	\end{equation}
	where 
	%To deal with the random component, consider for $a\in\lbrace0,1\rbrace$ the following classes of functions. 
	\begin{equation} \label{pol}
		\mathcal F_{n,a}=\left\{ g^{\tiny (a)}_{\tiny{\bbe},\xx_0}(\xx,y)=y^{\tiny a}
		K\left( \frac{\bbe (\xx_0 - \xx)}{h_n} \right) : \bbe \in \mathbb R^{p\times d } \hbox{ semi-orthogonal, } \xx_0\in\mathbb{R}^{p} \right\}.
	\end{equation}

\end{proposition}

\textbf{Proof of Proposition \ref{desig-max}.} 
Let
\begin{align*}
	\mathcal{F}_{a}=\left\lbrace f^{\tiny (a)}_{\tiny{\bbe},\mathbf{x}_0,s}(\xx,y)=y^{a} K \left(\frac{\bbe(\xx_0 - \xx)}{s}\right): \mathbf{x}_0 \in \mathbb{R}^{p}, \bbe \in \mathbb{R}^{p\times d }, 0<s  \right\rbrace \cup \lbrace 0 \rbrace.
\end{align*}
Note that
\begin{align*}
	\sup_{\mathcal{F}_{n,a}} \vert \PP_n f - Pf \vert \leq \sup_{\mathcal{F}_{a}} \vert P_n f - Pf \vert,
\end{align*}
since $\mathcal{F}_{n,a} \subseteq \mathcal{F}_{a}$.
We will show that for some constant $M_{a}>0$
\begin{align}\label{pol1}
	\EE \left[ \sup_{\mathcal{F}_{a}} \vert P_n f - Pf \vert \right] \leq M_{a} \frac{1}{ \sqrt{n}}, 
\end{align}
and, therefore, (\ref{random}) follows.
To prove (\ref{pol1}), we will apply the maximal inequalities of Theorem 4.2 from  \cite{pollard1989} to $\mathcal{F}_{a}$.

Let
\begin{align*}
	\mathcal{P}=\left\lbrace p_{\mathbf{x}_{0},\tiny{\bbe},s}(\xx)=\left\Vert \frac{\bbe (\mathbf{x}_0 - \mathbf{x})}{s}\right \Vert : \mathbf{x}_{0}\in \mathbb{R}^{p}, \bbe \in \mathbb{R}^{p\times d }    \hbox{ semi-orthogonal, }    s>0\right\rbrace.
\end{align*}
It is easy to show that $\mathcal{P}$ is a VC class of functions, since it can be obtained by applying a monotone function, the square root, to a subset of a vector space $s$  of polynomials; see Lemmas 2.6.15 and 2.6.18 (viii) of \cite{van1996}.
Since $k$ is of bounded variation, there exist bounded monotone functions $k_1$ and $k_2$ such that $k=k_1-k_2$. By Lemma 2.6.18 (viii) of van der Vaart and Wellner (1996), $k_1(\mathcal{P})$ and $k_2(\mathcal{P})$ are VC. Note that $k_1(\mathcal{P})$ and $k_2(\mathcal{P})$ have contant envelopes $K_1(\xx)=\Vert k_1 \Vert_{\infty}$ and $K_2(\xx)=\Vert k_2 \Vert_{\infty}$ respectively. By Theorem 2.6.7 of \cite{van1996}  there exists a constant $L_1>0$ such that for $i=1,2$, for all $0<\varepsilon<1$ and probability measures with finite support $Q$
\begin{align*}
	N(\varepsilon \Vert K_i \Vert_{2,Q},k_i(\mathcal{P}),\Vert . \Vert_{2,Q}) \leq L_1 \varepsilon^{-N_i}
\end{align*}
where $N_i$ is a positive integer.
Consider the class of functions $k_1(\mathcal{P}) - k_2(\mathcal{P})$. It has envelope $\tilde  K(\xx)=\Vert k_1 \Vert_{\infty} +\Vert k_2 \Vert_{\infty}$. As a consequence for some positive integer $N>0$ and fixed constant $L_2>0$
\begin{align}
	N(\varepsilon \Vert \tilde K \Vert_{2,Q},k_1(\mathcal{P}) - k_2(\mathcal{P}),\Vert . \Vert_{2,Q}) \leq L_2 \varepsilon^{-N}.
	\label{Eq:cota-cover}
\end{align}

%Now $\mathcal{F}_{a}\subset y^{a}(g_1(\mathcal{P}) - g_2(\mathcal{P})) \cup \lbrace 0 \rbrace$ . $\mathcal{F}_{a}$ has envelope $F_a(y,\mathbf{x})=G(\mathbf{x}) \vert y \vert^{a}$.
%It follows easily from \eqref{Eq:cota-cover} that for some constants $C_1,V>0$
%\begin{align*}
%D(\varepsilon \Vert F_{a} \Vert_{2,Q}, \mathcal{F}_{a},\Vert . \Vert_{2,Q})\leq C_1 \varepsilon^{-V}
%\end{align*}
%for all $0<\varepsilon<1$ and all probability measures with finite support $Q$ such that $\Vert F_{a} \Vert_{2,Q}>0$. 

Using that $\mathcal{F}_{a}\subset y^{a}(k_1(\mathcal{P}) - k_2(\mathcal{P})) \cup \lbrace 0 \rbrace$ and \eqref{Eq:cota-cover} it is easy to show that $\mathcal{F}_{a}$ is manageable in the sense of \cite{pollard1989} for the envelope $F_a(y,\mathbf{x})=\tilde K(\mathbf{x}) \vert y \vert^{a}$.
Hence, by Theorem 4.2 of \cite{pollard1989}, for some constant $M_{a}>0$,
\begin{align*}
	\EE \big[ \sup_{\mathcal{F}_{a}} \vert P_n f - Pf \vert \big] \leq  M_{a} \frac{1}{\sqrt{n}}.
\end{align*}

\begin{proposition}
	\label{desig-max2}
	Consider a kernel $K$ that  satisfies S.1. Let  $a\in \{0,1\}$. Assume that for some constant $C$,  $\vert Y \vert \leq C$ with probability one and $(h_n^{d}n)/ \log n \rightarrow \infty$.  Then
	\begin{equation}
		\frac{1}{h_{n}^{d}}\sup_{f\in \mathcal F_{n,a} } \vert \PP_n f - 
		\PP f\vert \rightarrow 0 \:\: a.s..
		\nonumber
	\end{equation}
\end{proposition}

\textbf{Proof of Proposition \ref{desig-max2}.}
Let
\begin{align*}
	\mathcal{H}_{n,a}=\lbrace (C\Vert K\Vert_{\infty})^{-1} f : \;\; f \in \mathcal{F}_{n,a}\rbrace.
\end{align*}
Since $Y$ is bounded by $C$, for all $l \in  \mathcal{H}_{n,a} $, $\vert l \vert \leq 1$ and
\begin{align*}
	P(l^2)\leq C_2 h_{n}^{d},
\end{align*}
for some fixed constant $C_2$.
Now, using the same arguments used in the proof of Proposition \ref{desig-max} it follows that for any probability measure $Q$ and $0<\varepsilon<1$
\begin{align*}
	N(\varepsilon ,\mathcal H_{n,a},\Vert . \Vert_{1,Q}) \leq C_1 \varepsilon^{-V}.
\end{align*}
for some positive constants $C_1$ and $V$.

Hence, applying Theorem 37 of \cite{pollard1984} to $\mathcal{H}_{n,a}$ with $\alpha_n=1$ and $\delta_n=C_2 h_{n}^{d/2}$ we have that
\begin{equation}
	\frac{1}{h_{n}^{d}}\sup_{h\in \mathcal H_{n,a} } \vert \PP_n h - 
	\PP h\vert \rightarrow 0 \:\: a.s.
	\nonumber
\end{equation}
which proves the result.

\textbf{Proof of Theorem \ref{consistency}}
We will now show that \eqref{cada_uno} holds. 
We will control the bias and the  random component, both for $a=0$ and for $a=1$.
\begin{eqnarray*}
	\label{bias-random}
	\sup_{\xx_0 \in \mathcal{K}} \left\vert    \frac{1}{h_n^d}\PP_{n} \gn^{(a)}-S_a(\bbe_0,\xx_0)\right\vert &\leq &
	\sup_{\xx_0 \in \mathcal{K}} \left \vert  \frac{1}{h_n^d} \PP_{n} \gn^{(a)}-S_a(\widehat\bbe_n,\xx_0,h_n)\right\vert \\
	&&+ \sup_{\xx_0 \in \mathcal{K}} 
	\left\vert S_a(\widehat\bbe_n,\xx_0,h_n) - S_a(\bbe_0,\xx_0) \right\vert 
\end{eqnarray*}
The second term in (\ref{bias-random}) converges to zero in probability as a consequence of Proposition \ref{bias}.
Since
\begin{eqnarray*}
	\left \vert  \frac{1}{h_n^d} \PP_{n} \gn^{(a)}-S_a(\widehat\bbe_n,\xx_0,h_n)\right\vert 
	&\le &  \frac{1}{h_n^d} \sup_{\mathcal{F}_{n,a}}\left \vert  \PP_{n} f - Pf \right\vert,
\end{eqnarray*}
when (i) holds the theorem follows from Proposition \ref{desig-max2}. When 
(ii) holds, by Markov's inequality  for any $\varepsilon>0$
\begin{align*}
	\mathbb{P}\left( \sup_{\xx_0 \in \mathcal{K}} \left \vert  \frac{1}{h_n^d}\PP_{n} \gn^{(a)}-S_a(\widehat\bbe_n,\xx_0,h_n)\right\vert\geq \varepsilon\right )
	&\leq
	\mathbb{P}  \left( \frac{1}{h_n^d} \sup_{\mathcal{F}_{n,a}}\left \vert  \PP_{n} f - Pf \right\vert\geq \varepsilon\right ) \\
	&\leq \frac{1}{\varepsilon}\frac{1}{h_n^d} \EE \sup_{\mathcal{F}_{n,a}}
	\left \vert  \PP_{n} f - Pf  \right \vert
\end{align*}
that converge to 0 by  Proposition \ref{desig-max} when $h_n^{2d} n \rightarrow \infty$.
%When (ii) holds
%\begin{align*}
% \mathbb{P}\left( \sup_{\xx_0 \in \mathcal{K}} \left \vert  \frac{1}{h_n^d}\PP_{n} \gn^{(a)}-S_a(\widehat\bbe_n,\xx_0,h_n)\right\vert\geq \varepsilon\right )
% \leq
%  \mathbb{P}  \left( \frac{1}{h_n^d} \sup_{\mathcal{F}_{n,a}}\left \vert  \PP_{n} f - Pf \right\vert\geq \varepsilon\right ) \rightarrow 0
%\end{align*}
%since 
%\begin{align*}
%\frac{1}{h_n^d} \sup_{\mathcal{F}_{n,a}}\left \vert  \PP_{n} f - Pf \right\vert \rightarrow 0 \:\: a.s.
%\end{align*}
%by Lemma \ref{desig-max2}
The theorem follows since
$\inf_{\xx_0 \in \mathcal{K}}\PP_{n} \gn^{(0)}\geq \inf_{\xx_0 \in \mathcal{K}}S_0(\bbe_0,\xx_0)-\sup_{\xx_0 \in \mathcal{K}}\vert\PP_{n} \gn^{(0)}-S_0(\bbe_0,\xx_0)\vert$.

		Finally, note that
				\begin{eqnarray*}
			\sup_{\mathbf{x}_0 \in \mathcal{K}} \vert \widehat\eta(\xx_0)-  \eta(\xx_0) \vert&\leq& 
			\left[\sup_{\xx_0 \in \mathcal{K}}\vert \frac{1}{h_n^d}\PP_{n} \gn^{(1)}-S_1(\bbe_0,\xx_0)\vert \vert S_0(\bbe_0,\xx_0)\vert\right.\\
&&	\left.	+\sup_{\xx_0 \in \mathcal{K}}\vert \frac{1}{h_n^d}\PP_{n} \gn^{(0)}-S_0(\bbe_0,\xx_0)\vert \vert S_1(\bbe_0,\xx_0)\vert \right]	\\\
&&\;\;\;\;\; \left[{\inf_{\xx_0 \in \mathcal{K}}\frac{1}{h_n^d}\PP_{n} \gn^{(0)}}S_0(\bbe_0,\xx_0)\right]^{-1}
		\end{eqnarray*}
Since   $\inf_{\xx_0 \in \mathcal{K}}\PP_{n} \gn^{(0)}\geq \inf_{\xx_0 \in \mathcal{K}}S_0(\bbe_0,\xx_0)-\sup_{\xx_0 \in \mathcal{K}}\vert\PP_{n} \gn^{(0)}-S_0(\bbe_0,\xx_0)\vert$, the result announced in Theorem \ref{consistency} follows from 
(\ref{cada_uno}) and assumption \ref{den_unif_bounded}.

\subsection{Proof of Proposition \ref{sequence}}

To prove Proposition  \ref{sequence} we need some previous  results.
We start with an algebraic one.
\begin{proposition}
	\label{al}

	(Linear Algebra) Let $\proy_n\in \mathbb R^{p\times p}$ be a sequence of	orthogonal projections of rank  $d$
	that converges to an orthogonal projections matrix $\proy_0$, which also has rank  $d$. Let $\bbe_0\in \mathbb R^{p\times d}$ be an orthonormal base of $\hbox{span}(\proy_0)$: 
	$\proy_0\bbe_0=\bbe_0$ and $\bbe_0^T\bbe_0=I_d$. 
	Then, for $\widetilde\bbe_n:=\proy_n \bbe_0$,  $\widetilde\bbe_n\to \bbe_0$ and,  for $n$ large enough, is a base of $span(\proy_n)$. 
	Moreover, defining   $\bbe_n=(\widetilde\bbe_n^\trans \widetilde\bbe_n)^{-1/2}\widetilde\bbe_n$ we get that $\bbe_n\to \bbe_0$, $\proy_n\bbe_n=\bbe_n$ and $\bbe_n^\trans\bbe_n=I_{d}$.
\end{proposition}

\textbf{Proof of Proposition \ref{al}: } 
The convergence of $\widetilde {\bbe}_n\to \bbe_0$ follows from that of $\proy_n\to\proy_0$.
To finish the proof, it is enough to show that $\widetilde \bbe_n$ is linearly independent. For that, assume   that there exists a sequence $n_k$ such that the columns of $\widetilde {\bbe}_{n_k}$ are not linearly independent. That means that there exists a sequence $a_{n_k}\in \mathbb R^ p$, with $\Vert a_{n_k}\Vert =1$ such that $\widetilde {\bbe}_{n_k}a_{n_k}=0$. We can then choose a subsequence, which  we also denote with ${n_k}$, such that $a_{n_k}\to a$, with $\Vert a\Vert =1$. Combining this fact with  the convergence of $\widetilde{\bbe}_n$ to $\widetilde{\bbe}_0$, 
we get that $\bbe_0 a=0$, which contradicts that $\bbe_0$ has rank $d$.

\begin{proposition}
	\label{invarianza}
	Assume that $\widehat \proy_n\in \mathbb R^{p\times p}$,  $\widehat\proy_n\convprob \proy_0$, with $\hbox{rank}(\widehat \proy_n)=d$,  $\hbox{rank}(\proy_0)=d$. 
	Fix $\bbe_0$
	such that
	$\proy_0\bbe_0^T=\bbe_0^T$ and $\bbe_0\bbe_0^T=I_d$, 
	define 
	$\widetilde\bbe_n:=\widehat\proy_n \bbe_0^T$. For each $n$  consider the event 
	\begin{equation*}
		\label{An}
		\mathcal A_n:=\left\{\widetilde  \bbe_n^T\;\hbox{is a basis of $span(\widehat \proy_n)$}\right\}.
	\end{equation*}
	Then,  $\mathbb P(\mathcal A_n)\to 1$.	Also, on $\mathcal A_n$ we can define $\widehat \bbe_n=(\widetilde\bbe_n \widetilde\bbe_n^T)^{-1/2}\widetilde\bbe_n$
	and get that   
	$\widehat {\bbe}_n\convprob \bbe_0\;,\widehat\proy_n\widehat {\bbe}_n^T=\widehat {\bbe}_n^T\;\hbox{and}\;\widehat {\bbe}_n\widehat {\bbe}_n^T=I_{d}.$
	Moreover, if $\sqrt{n}(\widehat\proy_n-\proy_0)=O_P(1)$, then $\sqrt{n}(\widehat{\bbe}_n-\bbe_0)=O_P(1)$.
\end{proposition}

\textbf{Proof of Proposition  \ref{invarianza}:}
Assume that there exists $\delta>0$ and a subsequence $n_k$ such that 
\begin{equation}
	\label{malos}
	\mathbb P(\mathcal A_{n_k})\leq 1-\delta.
\end{equation}
From the convergence in probability of $\widehat \proy_n$ to $\proy_0$, there exists a further subsequence for which the convergence holds a.e. For sake of simplicity, we keep using $n_k$ for such a subsequence. Let $\mathcal A$ denote the set of total probability where  $\widehat \proy_n$ to $\proy_0$. 
Thus, 
$$\mathcal A\subseteq \displaystyle{\cup_m \cap_{n_k\geq m} \mathcal A_{n_k}}, $$
and so 
$$1=\PP(\mathcal A)\leq \displaystyle{\lim_{m\to \infty}\PP\left(\cap_{n_k\geq m} \mathcal A_{n_k}\right)} \leq \displaystyle{\lim_{n_k\to \infty} \PP\left(\mathcal A_{n_k}\right)}, 
$$
contradicting (\ref{malos}).

\textbf{Proof of Proposition  \ref{sequence}:}
The sequence $\widehat \bbe_n$ constructed in Proposition \ref{invarianza} satisfies the conditions required.

\bigskip

%\bibliographystyle{apalike}

%\bibliography{refs}

\end{document}